\documentclass[sigconf]{acmart}

\usepackage{booktabs} 
\usepackage{IEEEtrantools}
\usepackage{latexsym,amsfonts,amsmath} 
\usepackage{graphicx}
\usepackage{svg}
\usepackage{dsfont}
\usepackage{enumerate}
\usepackage{hyperref}

\newtheorem{algorithm}{Algorithm}

\newtheorem{remark}{Remark}

\usepackage{xcolor}

\usepackage{subfig}

\usepackage{listings}
\definecolor{codegreen}{rgb}{0,0.6,0}
\definecolor{codegray}{rgb}{0.5,0.5,0.5}
\definecolor{codepurple}{rgb}{0.58,0,0.82}
\definecolor{backcolour}{rgb}{0.9529411765, 0.9568627451, 0.9647058824}
\lstdefinestyle{mystyle}{
    basicstyle=\ttfamily\footnotesize,
    backgroundcolor=\color{backcolour},   
    commentstyle=\color{codegreen},
    keywordstyle=\color{magenta},
    numberstyle=\tiny\color{codegray},
    stringstyle=\color{codepurple},
    basicstyle=\ttfamily\footnotesize,
    breakatwhitespace=false,         
    breaklines=true,                 
    captionpos=b,       
    keepspaces=true,                 
    numbers=left,       
    xleftmargin=10pt,
    numbersep=5pt,                  
    showspaces=false,                
    showstringspaces=false,
    showtabs=false,                  
    tabsize=2
}
\usepackage{mathrsfs}

\usepackage{algorithm}
\usepackage{algpseudocode}
\usepackage{xspace}

\definecolor{purple}{HTML}{9333EA}

\newcommand{\identity}{\mathbb{I}}

\newcommand{\initial}{X_I}
\newcommand{\obstruction}{X_O}

\begin{document}

\title[TRUST: Controller Synthesis for Unknown Models Using a Single Trajectory]{TRUST: Stabili\underline{\textbf{T}}y and Safety Cont\underline{\textbf{R}}oller Synthesis for  \underline{\textbf{U}}nknown Dynamical Models Using a \underline{\textbf{S}}ingle \underline{\textbf{T}}rajectory}

\author{Jamie Gardner}
\affiliation{%
	\institution{School of Computing}
	\institution{Newcastle University}
		\country{}}
\email{j.gardner3@newcastle.ac.uk}

\author{Ben Wooding}
\affiliation{%
	\institution{School of Computing}
	\institution{Newcastle University}
		\country{}}
\email{ben.wooding@newcastle.ac.uk}

\author{Amy Nejati}
\affiliation{%
	\institution{School of Computing}
	\institution{Newcastle University}
		\country{}}
\email{amy.nejati@newcastle.ac.uk}

\author{Abolfazl Lavaei}
\affiliation{%
	\institution{School of Computing}
	\institution{Newcastle University}
		\country{}}
\email{abolfazl.lavaei@newcastle.ac.uk}

\thanks{Jamie Gardner and Ben Wooding have contributed equally.}

\begin{abstract}
\textsf{TRUST} is an open-source software tool developed for \emph{data-driven controller synthesis} of dynamical systems with \emph{unknown} mathematical models, ensuring either stability or safety properties. By collecting only a \emph{single input-state trajectory} from the unknown system and satisfying a rank condition that ensures the system is  \textsf{persistently excited} according to the Willems \textit{et al.}'s fundamental lemma, \textsf{TRUST} aims to design either \emph{control Lyapunov functions (CLF)} or \emph{control barrier certificates (CBC)}, along with their corresponding stability or safety controllers. The tool implements sum-of-squares (SOS) optimization programs solely based on data to enforce stability or safety properties across four system classes: \emph{(i) continuous-time nonlinear polynomial systems, (ii) continuous-time linear systems, (iii) discrete-time nonlinear polynomial systems, and (iv) discrete-time linear systems.} \textsf{TRUST} is a Python-based web application featuring an intuitive, reactive graphic user interface (GUI) built with web technologies. It can be accessed at \texttt{https://trust.tgo.dev} or installed locally, and supports both manual data entry and data file uploads. Leveraging the power of the Python backend and a JavaScript frontend, \textsf{TRUST} is designed to be highly user-friendly and accessible across desktop, laptop, tablet, and mobile devices. We apply \textsf{TRUST} to a set of physical benchmarks with unknown dynamics, ensuring either stability or safety properties across the four supported classes of models.
\end{abstract}


\maketitle

\section{Introduction}\label{sec:introduction}

The formal synthesis of controllers that ensure stability or safety of dynamical systems is a cornerstone of control theory, especially in safety-critical applications where failures can lead to loss of life or significant financial consequences~\cite{mcgregor2017analysis}. These applications include robotics, aerospace, autonomous vehicles, and medical devices, where system reliability is paramount. Traditionally, control synthesis has relied heavily on model-based approaches, which require precise mathematical representations of system dynamics. However, obtaining models with closed-form expressions of their dynamic can be highly challenging, as the identified models may not fully capture the complexities of real-world systems~\cite{hou2013model}. In response to this key obstacle, \emph{direct} data-driven techniques have emerged in the literature as a compelling alternative~\cite{dorfler2022bridging}. These techniques leverage system data to directly design controllers without the need for an explicit mathematical model, offering a more flexible and practical approach for complex dynamical systems.

While control Lyapunov functions (CLF) are essential methods for ensuring stability in dynamical systems~\cite{khalil2002nonlinear}, \emph{control barrier certificates (CBC)} have been introduced as a powerful method for guaranteeing safety \citep{prajna2004safety,ames2019control,xiao2023safe}. Similar to Lyapunov functions, CBCs are defined over the system's state space, but they focus on satisfying specific inequalities on both the system dynamics and the function itself. By identifying an appropriate \emph{level set} of CBC, unsafe regions can be separated from the system's trajectories, starting from a given set of initial conditions. As a result, the existence of such a function not only provides formal safety certification but also facilitates the design of a controller that enforces safety throughout the system's operation.

\subsection{Data-Driven Techniques}

Since constructing CLF or CBC typically requires precise mathematical models of the system dynamics, the development of data-driven techniques has become crucial in control theory. Two promising approaches in the literature offer formal controller synthesis  using collected data without relying on explicit models. The first is the \emph{scenario approach}~\citep{calafiore2006scenario,campi2009scenario}, which solves the problem by leveraging collected data and then translating the results back to unknown models using intermediate steps involving \emph{chance constraints}~\citep{esfahani2014performance}. While this method shows significant potential for providing formal guarantees for  systems with unknown models, it requires the data to be independent and identically distributed (i.i.d.). This restriction means that only one input-output data pair can be collected from each trajectory~\citep{calafiore2006scenario}, necessitating the collection of \emph{multiple independent trajectories}—potentially up to millions in real-world scenarios—to achieve a desired confidence level, based on a well-defined closed-form relationship between them.

An alternative to the scenario method is the \emph{non-i.i.d. trajectory-based approach}, which complements the former by requiring only a \emph{single input-output trajectory} from the unknown system over a specific time horizon~\citep{de2019formulas}. This approach utilizes the concept of \emph{persistent excitation}, where the trajectory should meet a rank condition to sufficiently excite the system’s dynamics, as outlined by Willems \textit{et al.}'s fundamental lemma~\citep{willems2005note}. Specifically, when a system is persistently excited, the trajectory provides enough information about the system’s behavior to enable its analysis. This method is advantageous since it eliminates the need for multiple independent trajectories, making it more practical in cases where obtaining several distinct trajectories is challenging.

\subsection{Data-Driven Literature}\label{subsec:related-literature}

A substantial body of research has investigated data-driven analysis of unknown systems using the scenario approach. This includes stability analysis of unknown systems (\emph{e.g.,}~\cite{kenanian2019data,boffi2021learning,lavaei2022data11,lavaei2023ISS}), the construction of barrier certificates (\emph{e.g.,}~\citep{nejati2023formal,lindemann2021learning,nejati2023data,aminzadeh2024physics}) and finite abstractions (\emph{e.g.,}~\citep{kazemi2024data,makdesi2023data,devonport2021symbolic,coppola2022data,ajeleye2023data}), both applied in safety verification and controller synthesis. Additionally, there is a rich literature on data-driven methods based on the \emph{single-trajectory approach}. These approaches have been used to ensure stability and invariance properties in unknown systems (\emph{e.g.,}~\citep{berberich2020data,dai2023data,luppi2024data,zaker2024certified}) and to construct control barrier certificates for both continuous- and discrete-time systems (\emph{e.g.,}~\citep{nejati2022data,samari2024singletrajectory,wooding2024learning}).

\subsection{Related Software Tools}\label{subsec:related-software}
There are only a few software tools available for constructing either control Lyapunov functions or (control) barrier certificates. One such tool is the class library developed in~\cite{bjornsson2015class}, which computes Lyapunov functions for nonlinear systems in C++. Another tool, LyZNet~\cite{liu2024tool}, utilizes neural networks (NN) to learn and verify stability functions and regions of attraction, leveraging a physical model to inform the NN. The tool FOSSIL 2.0~\cite{edwards2024fossil} synthesizes stability and safety barrier functions for both discrete and continuous systems using a model-driven NN. PRoTECT~\cite{wooding2024protect} is another tool that designs safety barrier certificates for discrete- and continuous-time systems. While these tools~\cite{bjornsson2015class,liu2024tool,edwards2024fossil,wooding2024protect} demonstrate significant potential, they all require \emph{precise mathematical models of dynamical systems} to construct either stability Lyapunov functions or safety barrier certificates, along with their corresponding controllers. This requirement potentially contrasts with real-world scenarios where precise system models are often unavailable, making the aforementioned tools impractical for those applications.

\subsection{Central Contributions}\label{subsec:contributions}
Motivated by the central challenge of unknown system models, this tool paper introduces the following innovative contributions:
\begin{enumerate}[(i)]
\item \textsf{TRUST} is a first-of-its-kind tool that leverages \emph{data-driven techniques} to synthesize stability Lyapunov functions and safety barrier certificates, along with their corresponding controllers, using only a \emph{single input-state trajectory} from unknown systems.
\item \textsf{TRUST} leverages the sum of squares (SOS) optimization toolbox~\cite{prajna2002introducing}, powered by MOSEK\footnote{https://www.mosek.com/documentation/}, and supports \emph{four classes of dynamical systems}: (i) continuous-time nonlinear polynomial systems (ct-NPS), (ii) continuous-time linear systems (ct-LS), (iii) discrete-time nonlinear polynomial systems (dt-NPS), and (iv) discrete-time linear systems (dt-LS).
\item Implemented as a responsive and reactive Python Flask\footnote{https://flask.palletsprojects.com/en/stable/} web application, \textsf{TRUST} offers an intuitive, user-friendly interface, allowing users to seamlessly interact with the tool. Users can directly access and work with the application through the web without the need for downloads or installations.
\item \textsf{TRUST} also offers a \emph{Docker-based version} that can be downloaded and run locally, potentially achieving higher speeds depending on the capabilities of the local machine.
\item The server-side Python application, built on the Flask web framework, follows the Model-View-Controller (MVC)~\cite{bucanek2009model} architecture
and adheres to Test-Driven Development (TDD)~\cite{beck2022test} practices, ensuring high-quality and maintainable code.
\item \textsf{TRUST}
leverages a monolithic architecture, using InertiaJS\footnote{https://inertiajs.com/}
to enable real-time client-side updates and effective error handling, ensuring a seamless user experience across all platforms without the need to build a separate Application Programming Interface (API).
\item The tool supports both manual data entry and data file uploads for collected trajectories, offering a user-friendly GUI for inputting the required information, such as the state space, and initial and multiple unsafe sets for designing CBC.
\item We demonstrate the effectiveness of \textsf{TRUST} through a series of physical benchmarks, covering the four classes of dynamical systems and showcasing their respective stability or safety properties.
\end{enumerate}
The web-based version of \textsf{TRUST} is accessible at:
 \begin{center}
 	\href{https://trust.tgo.dev}{https://trust.tgo.dev}
 \end{center}
The \textsf{TRUST} source code, accompanied by comprehensive installation and usage instructions for the Docker-based version, can be accessed at:
\begin{center}
    \href{https://github.com/Kiguli/TRUST}{https://github.com/Kiguli/TRUST}
\end{center}

\subsection{Notation}\label{subsec:notation}
The symbols $\mathbb{R}$, $\mathbb{R}^+_0$, and $\mathbb{R}^+$ represent the sets of real numbers, non-negative and positive real numbers, respectively. Similarly, the symbols $\mathbb{N}$ and $\mathbb{N}^+$ denote the sets of natural numbers, including and excluding zero, respectively.
The notation $\mathbb{R}^{n \times m}$ denotes a matrix of size $n \times m$ with real values, while $\mathbb{R}^n$ represents a vector of size $n$. The symbols $A^{-1}$ and $A^\top$ represent the inverse and the transpose of matrix $A \in \mathbb R^{n\times n}$, respectively. The notation $[a,b]$ refers to the closed interval between $a$ and $b$. The identity matrix in $\mathbb R^{n\times n}$ is denote by $\identity_n$. We denote a \emph{symmetric} positive-definite matrix $P \in \mathbb R^{n\times n}$ as $P \succ 0$.

\section{Overview of \textsf{TRUST}}\label{sec:overview-of-the-tool}

{\bf Modes of the Tool.} \textsf{TRUST} is capable of solving two main types of control problems:
\begin{enumerate}[(i)]
\item {\bf Stability problems} –  synthesizing a control Lyapunov function and its corresponding controller to enforce the stability of the unknown system. The stability property implies that as time approaches infinity, the system's trajectories converge to the origin, which serves as the equilibrium point.
\item {\bf Safety problems} – synthesizing a control barrier certificate and its corresponding controllers to ensure the safe behavior of the unknown system. The safety property ensures that the system's trajectories, starting from an initial region, do not reach (potentially multiple) unsafe regions within an infinite time horizon.
\end{enumerate}

\noindent {\bf Classes of Systems.} \textsf{TRUST} can support \emph{four classes of systems}: (i) continuous-time nonlinear polynomial systems (ct-NPS), (ii) continuous-time linear systems (ct-LS), (iii) discrete-time nonlinear polynomial systems (dt-NPS), and (iv) discrete-time linear systems (dt-LS). For each of these system classes, \textsf{TRUST} can synthesize both CLF and CBC, along with their respective stability and safety controllers.

\noindent {\bf Datasets (\texttt{*.csv}, \texttt{*.txt}, \texttt{*.json}).}
\textsf{TRUST} accepts input datasets that include a \emph{persistently excited} data trajectory, which should be provided in one of the following common file formats: \texttt{*.csv}, \texttt{*.txt}, or \texttt{*.json}.  For discrete-time systems, the data should be collected over a time horizon $[0,1,\ldots,T-1]$, where $T\in\mathbb N^+$ is the number of collected samples:
\begin{subequations}\label{eq:data-discrete}
\begin{align}
\mathcal{U}_{0}^d &= [u(0),u(1),u(2),\ldots,u(T-1)] &\in\mathbb R^{m\times T},\label{New1}\\
\mathcal{X}_{0}^d  &= [x(0),x(1),x(2),\ldots,x(T-1)] &\in\mathbb R^{n\times T},\\
\mathcal{X}_{1}^d  &= [x(1),x(2),x(3),\ldots,x(T)] &\in\mathbb R^{n\times T},\label{New2}\\
\mathcal{N}_{0}^d  &= [\mathcal{M}(x(0)), \mathcal{M}(x(1)),\ldots, \mathcal{M}(x(T-1))] &\in\mathbb R^{N\times T},\label{New3}
\end{align}
\end{subequations}
where $x\in \mathbb R^n$ and $u\in \mathbb R^m$ are the state and input variables of the system. Moreover, $\mathcal{M}(x)\in\mathbb{R}^N$ is a vector of monomials in state $x\in \mathbb R^n$, which shapes the dynamics of nonlinear polynomial systems. We call the collected data in \eqref{New1}-\eqref{New2} a \emph{single input-state trajectory}. Note that the trajectory $\mathcal{N}_{0}^d$ in \eqref{New3} is constructed from the trajectory $\mathcal{X}_{0}^d$ and the form of the monomial vector $\mathcal{M}(x)\in\mathbb{R}^N$.

Similarly, the \emph{continuous-time data} can be collected over the time interval $[t_0 , t_0 + (T - 1)\tau]$, with $T \in \mathbb N^+$ being the number of collected samples, and $\tau \in \mathbb R^+$ as the sampling time:
\begin{subequations}\label{eq:data-continuous}
\begin{align}\label{new5}
	\mathcal U_{0}^c  &\!=\! [u(t_0), u(t_0 + \tau), \dots, u(t_0 + (T - 1)\tau)] &\in\mathbb R^{m\times T},\\\label{new6}
	\mathcal X_{0}^c &\!=\! [x(t_0), x(t_0 + \tau), \dots, x(t_0 + (T - 1)\tau)] &\in\mathbb R^{n\times T},\\\label{new7}
	\mathcal X_{1}^c &\!=\! [\dot x(t_0), \dot x(t_0 + \tau), \dots, \dot x(t_0 + (T - 1)\tau)] &\in\mathbb R^{n\times T},\\\label{new8}
    \mathcal N_{0}^c &\!=\! [\mathcal M(x(t_0))~~\dots~~\mathcal M(x(t_0 + (T - 1)\tau))] &\in\mathbb R^{N\times T}.
\end{align}
\end{subequations}

To establish the theoretical aspects of the single-trajectory approach, it is required that the trajectories $\mathcal{N}_{0}^c$ and $\mathcal{N}_{0}^d$ for nonlinear polynomial cases, and $\mathcal{X}_{0}^c$ and $\mathcal{X}_{0}^d$ for linear cases, have \emph{full row-rank} (cf. Lemmas \ref{lem:Q-ctNPS}, \ref{lem:Q-ctLS}, \ref{lem:Q-dtNPS} for different cases). To ensure this, the number of samples $T$ must be greater than $N$ in nonlinear cases and $n$ in linear scenarios. Given that $\mathcal{N}_{0}^c$, $\mathcal{N}_{0}^d$, $\mathcal{X}_{0}^c$ and $\mathcal{X}_{0}^d$ are all derived from sampled data, this condition can be readily verified during data collection. This approach avoids explicitly identifying the system, which would otherwise require $\begin{bmatrix}
	\mathcal{U}^d_0\\\mathcal{X}^d_0
\end{bmatrix}$ (for discrete-time linear systems) to be full row-rank, see~\cite{de2023learning2}.

\textsf{TRUST} \emph{automatically verifies} that the collected data satisfies the following rank condition, ensuring that the data is persistently excited~\cite{willems2005note,de2019formulas}.  In case the collected data does not fulfill the required full row-rank condition, the tool will return an error to the user stating:
\emph{``Error: The collected data must be full row-rank''}.

\noindent {\bf Outputs.} The outputs of \textsf{TRUST} depend on the selected mode. For stability problems, \textsf{TRUST} aims to return a CLF $\mathcal{V}(x)$ along with the corresponding stability controller, both derived from the collected data. For safety problems, it provides a CBC $\mathcal{B}(x)$, along with its initial and unsafe level sets, $\gamma$ and $\lambda$, as well as the corresponding safety controller, all based on the data. If the tool is unable to synthesize either, \textsf{TRUST} will return an error message and guide the user on potential issues.

\section{Data-Driven Results for Continuous-Time Systems}\label{sec:scenarios}

Here, we present the corresponding results for the data-driven safety and stability controller synthesis of continuous-time systems with both nonlinear polynomial and linear dynamics. Since our work is a tool paper, we aim to focus on the features and implementation of the software, while we refer to the existing literature for the technical details of the data-driven approach. We remind the reader that the collected data used in this section follows the form described in~\eqref{eq:data-continuous}.

\subsection{Safety and Stability of ct-NPS}\label{subsec:continuous-time-nonlinear-polynomial-systems}
We consider continuous-time nonlinear polynomial systems defined as follows.

\begin{definition}
\label{def:system-description-ct-NPS} [{\bf ct-NPS}]
A continuous-time nonlinear polynomial system (ct-NPS) is described by
\begin{equation}
\label{eq:ct-NPS}
\Sigma^c\!: \dot{x} = A\mathcal{M}(x)+Bu,
\end{equation}
where $A\in\mathbb{R}^{n\times N}$ and $B\in\mathbb{R}^{n\times m}$ are \emph{unknown} system and control matrices, $\mathcal{M}(x)\in\mathbb{R}^N$ is a vector of monomials in state $x\in X$, and $u\in U$ is a control input, with $X\subset\mathbb{R}^n$, and $U\subset\mathbb{R}^m$ being the state and input sets, respectively.
\end{definition}

The following lemma, taken from~\cite{guo2020learning,nejati2022data}, offers a data-based representation of closed-loop ct-NPS~\eqref{def:system-description-ct-NPS} with a controller $u=K(x)\mathcal{M}(x)$, where $K(x)\in\mathbb{R}^{m\times N}$ is a matrix polynomial.

\begin{lemma}[{\bf Data-based Representation of ct-NPS \cite{guo2020learning,nejati2022data}}]
\label{lem:Q-ctNPS}
Let $Q(x)$ be a $(T\times N)$ matrix polynomial such that
\begin{equation}
\label{eq:Q-matrix-ctNPS}
\mathbb{I}_N = \mathcal{N}_{0}^cQ(x),
\end{equation}
with $\mathcal N_{0}^c $ as in~\eqref{new8} being an $(N\times T)$ full row-rank matrix.
If one synthesize $u=K(x)\mathcal{M}(x) = \mathcal{U}_{0}^cQ(x)\mathcal{M}(x)$, then the closed-loop system $\dot{x}=A\mathcal{M}(x)+Bu$ has the following data-based representation:
\begin{equation}
\label{eq:A+BK=X1TQ-ctNPS}
\dot{x} = \mathcal{X}_{1}^cQ(x)\mathcal{M}(x),\quad\text{equivalently,}\quad A+BK(x) = \mathcal{X}_{1}^cQ(x).
\end{equation}
\end{lemma}

Using the above lemma, one can obtain a data-driven representation of ct-NPS with unknown matrices $A$ and $B$ as $\mathcal{X}_{1}^cQ(x)\mathcal{M}(x)$, which will be employed to design controllers for both CBC and CLF in the following theorems. It is worth noting that  to ensure $\mathcal{N}_{0}^c$ has full row rank, the number of samples $T$ must exceed $N$~\cite{guo2020learning}, a condition that can be readily fulfilled during data collection.

The following theorem, borrowed from~\cite[Theorem 8]{nejati2022data}, leverages the data-driven representation of ct-NPS from Lemma~\ref{lem:Q-ctNPS} to design a CBC $\mathcal B(x) = \mathcal M(x)^\top P \mathcal M(x)$,  with $P\in\mathbb{R}^{N\times N} \succ 0$,  and a safety controller $u= \mathcal{U}_{0}^cQ(x)\mathcal{M}(x)$ based on the collected data.

\begin{theorem}[{\bf Data-Driven CBC for ct-NPS~\cite{nejati2022data}}]
\label{thm:data-ctNPS-CBC}
Consider the ct-NPS in \eqref{eq:ct-NPS} with unknown matrices $A$, $B$, and its data-based representation $\dot{x}=\mathcal{X}_{1}^cQ(x)\mathcal{M}(x)$. Let $X_I, X_O \subset X$
represent the initial and unsafe regions of the ct-NPS, respectively. Suppose there exist constants $\gamma,\lambda \in \mathbb R^+$, with $\lambda > \gamma$, and a  matrix polynomial $H(x)\in\mathbb{R}^{T\times N}$ such that the following constraints are fulfilled:
\begin{subequations}
\begin{align}
\label{eq:ctNPS_CBC_data0}
	\mathcal{N}_{0}^cH(x)=P^{-1}&, \quad\quad\text{with} ~P \succ 0,\\
\label{eq:ctNPS_CBC_data1}
\mathcal{M}(x)^\top [\mathcal{N}_{0}^cH(x)]^{-1}\mathcal{M}(x) \leq \gamma, &\quad\quad\quad\forall x\in X_I, \\
\label{eq:ctNPS_CBC_data2}
\mathcal{M}(x)^\top [\mathcal{N}_{0}^cH(x)]^{-1}\mathcal{M}(x) \geq \lambda, &\quad\quad\quad\forall x\in X_O,\\
\label{eq:ctNPS_CBC_data3}
-[\frac{\partial \mathcal M}{\partial x}\mathcal{X}_{1}^cH(x) + H(x)^\top\mathcal{X}_{1}^{c\top}(\frac{\partial \mathcal M}{\partial x})^\top] &\geq 0,\quad \forall x\in X.
\end{align}
\end{subequations}
Then $\mathcal{B}(x) = \mathcal{M}(x)^\top [\mathcal{N}_{0}^cH(x)]^{-1}\mathcal{M}(x)$ is a CBC and $u = \mathcal{U}_{0}^cH(x)$\\$[\mathcal{N}_{0}^cH(x)]^{-1}\mathcal{M}(x)$ is its corresponding safety controller for the unknown ct-NPS, ensuring that any trajectory of the ct-NPS starting from $X_I$ will not reach $X_O$ within an infinite time horizon.
\end{theorem}

\begin{algorithm}[t]
	\caption{Data-driven design of \emph{CBC and safety} controller for \emph{ct-NPS}}\label{alg:ct-nonlinear}
	\begin{algorithmic}[1]
		\Require Regions of interest $X,X_I,X_O$, collected trajectories $\mathcal{U}_0^c,\mathcal{X}_0^c,\mathcal{X}_1^c$, a choice of monomials $\mathcal{M}(x)$\footnotemark[1]
		\State Check that the full row-rank condition for $\mathcal{N}_0^c$ is satisfied
		\State Solve~\eqref{eq:ctNPS_CBC_data0} and~\eqref{eq:ct-SOS3} for $P$ and $H(x)$ simultaneously\footnotemark[2]
		\State Given the constructed $H(x)$, solve \eqref{eq:ct-SOS1} and \eqref {eq:ct-SOS2} to design level sets $\gamma$ and $\lambda$, where $\lambda >\gamma$
		\Ensure CBC $\mathcal{B}(x) = \mathcal{M}(x)^\top [\mathcal{N}_{0}^cH(x)]^{-1}\mathcal{M}(x)$ and its corresponding safety controller $u = \mathcal{U}_{0}^cH(x)[\mathcal{N}_{0}^cH(x)]^{-1}\mathcal{M}(x)$
	\end{algorithmic}
\end{algorithm}

\footnotetext[1]{The selection of $\mathcal{M}(x)$ is the choice of the user. Based on the foundational work \cite{de2019formulas}, if an upper bound on the maximum degree of $\mathcal{M}(x)$ can be inferred using physical insights about the unknown system, $\mathcal{M}(x)$ should be chosen to encompass all possible combinations of states up to that upper bound. This ensures that $\mathcal{M}(x)$ potentially contains all monomial terms present in the actual unknown system.}

\footnotetext[2]{To satisfy condition~\eqref{eq:ctNPS_CBC_data0}, we define ${Z} = P^{-1}$ and enforce that it is a \emph{symmetric positive-definite} matrix, \emph{i.e.,} ${Z} \succ 0$. Once condition~\eqref{eq:ctNPS_CBC_data0} is met and ${Z}$ is designed, the matrix $P$ is computed as the inverse of ${Z}$, \emph{i.e.,} $P = {Z}^{-1}$.}

The following lemma, borrowed from~\cite[Corollary 11]{nejati2022data}, reformulates conditions~\eqref{eq:ctNPS_CBC_data0}-\eqref{eq:ctNPS_CBC_data3} as sum-of-squares (SOS) optimization programs, assuming that the regions of interest, $X$, $X_I$, and $X_O$, are semi-algebraic~\citep{bochnak2013real}. Specifically, a semi-algebraic set $X \subseteq \mathbb{R}^n$ is described by a vector of polynomials $a(x)$, meaning $X = \{x \in \mathbb{R}^n \mid a(x) \geq 0\}$, with the inequalities applied element-wise.

\begin{lemma}[{\bf Data-Driven SOS Reformulation of CBC for ct-NPS~\cite{nejati2022data}}]\label{Lemma_ct-NPS}
Consider the state set $X$, the initial set $X_I$, and the unsafe set $X_O$ as semi-algebraic sets, each defined by vectors of polynomial inequalities $g(x)$, $g_I(x)$, and $g_O(x)$, respectively. Suppose there exist constants $\gamma, \lambda \in \mathbb{R}^+$, with $\lambda > \gamma$, a matrix polynomial $H \in \mathbb{R}^{T \times N}$, and vectors of SOS polynomials $L_I(x)$, $L_O(x)$, and $L(x)$ such that the following conditions
\begin{subequations}
\begin{align}
	\label{eq:ct-SOS1}
	-\mathcal{M}(x)^\top [\mathcal{N}_{0}^cH(x)]^{-1}\mathcal{M}(x) -& L_I^\top(x)g_I(x) + \gamma, \\
	\label{eq:ct-SOS2}
	\mathcal{M}(x)^\top [\mathcal{N}_{0}^cH(x)]^{-1}\mathcal{M}(x) -& L_O^\top(x)g_O(x) - \lambda, \\
	\label{eq:ct-SOS3}
	-[\frac{\partial \mathcal M}{\partial x}\mathcal{X}_{1}^cH(x) + H(x)^\top\mathcal{X}_{1}^{c\top}(\frac{\partial \mathcal M}{\partial x})^\top] -& L^\top(x)g(x).
\end{align}
\end{subequations}
are all SOS polynomials, while condition~\eqref{eq:ctNPS_CBC_data0} is also fulfilled. Then, $\mathcal{B}(x) = \mathcal{M}(x)^\top [\mathcal{N}{0}^cH(x)]^{-1}\mathcal{M}(x)$ is a CBC, and $u = \mathcal{U}{0}^c H(x)$\\$[\mathcal{N}_{0}^c H(x)]^{-1} \mathcal{M}(x)$ is its corresponding safety controller for the unknown ct-NPS.
\end{lemma}

\begin{remark}
To accommodate an \emph{arbitrary number} of unsafe regions $X_{{O}_i}$, where $i \in \{1, \ldots, z\}$, condition \eqref{eq:ct-SOS2} should be repeated and enforced for each distinct unsafe region, a capability fully supported by \textsf{TRUST}.
\end{remark}

Algorithm~\ref{alg:ct-nonlinear} outlines the necessary steps for designing a CBC and its corresponding safety controller based solely on data.

The following theorem, borrowed from~\cite[Theorem 1]{guo2020learning}, provides the required conditions for designing a CLF $\mathcal{V}(x)=\mathcal{M}(x)^\top P\mathcal{M}(x)$, with $P \succ 0$, along with a stability controller $u= \mathcal{U}_{0}^cQ(x)\mathcal{M}(x)$ based on collected data.

\begin{theorem}[{\bf Data-Driven CLF for ct-NPS \cite{guo2020learning}}]\label{thm:ctNPS-stable}
Consider the ct-NPS in \eqref{eq:ct-NPS} with unknown matrices $A$, $B$, and its data-based representation $\dot{x}=\mathcal{X}_{1}^cQ(x)\mathcal{M}(x)$. Suppose there exists a polynomial matrix $H(x)\in\mathbb{R}^{T\times N}$ such that the following constrains are satisfied:
\begin{subequations}
\begin{align}
\label{eq:stability-ctNPS0}
	\mathcal{N}_{0}^cH(x)=P^{-1}&, \quad\quad\text{with} ~P \succ 0,\
\\\label{eq:stability-ctNPS1}
-[\frac{\partial \mathcal M}{\partial x}\mathcal{X}_{1}^cH(x) + H(x)^\top\mathcal{X}_{1}^{c\top}&(\frac{\partial \mathcal M}{\partial x})^\top] \succ 0
\end{align}
\end{subequations}
Then, $\mathcal{V}(x) = \mathcal{M}(x)^\top [\mathcal{N}_{0}^cH(x)]^{-1}\mathcal{M}(x)$ is a CLF, and $u = \mathcal{U}_{0}^cH(x)$\\$[\mathcal{N}_{0}^cH(x)]^{-1}\mathcal{M}(x)$ is its corresponding stability controller for the unknown ct-NPS, ensuring that any trajectory of the ct-NPS converges to the origin as its equilibrium point when time approaches infinity.
\end{theorem}

\begin{figure*}[h!]
	\includegraphics[
	width=1\textwidth
	]{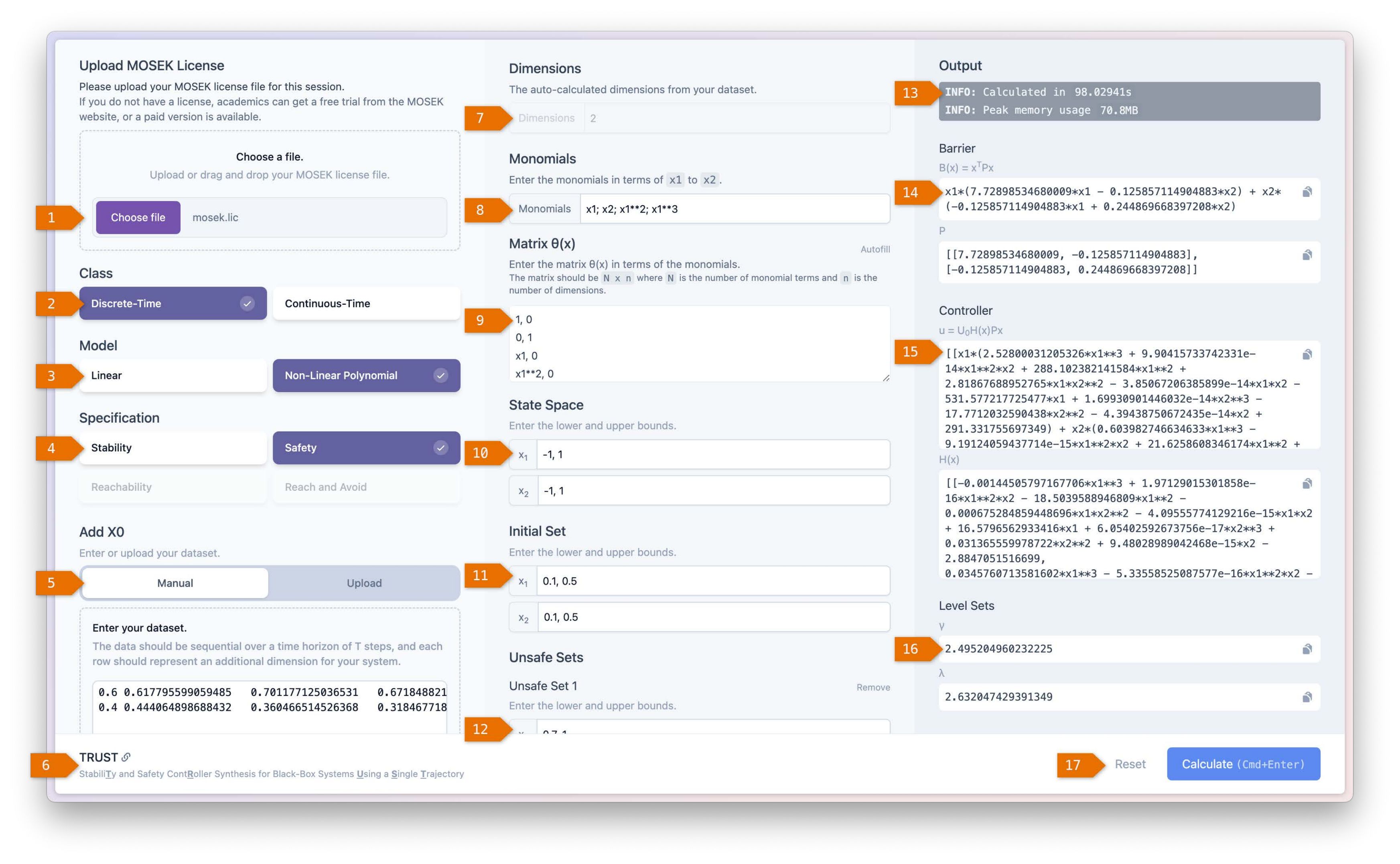}\vspace{-0.7cm}
	\caption{\textsf{\textsf{TRUST} GUI with numbered annotations indicating respective sections.}}
	\label{fig:TRUST}
\end{figure*}

\begin{algorithm}[t!]
	\caption{Data-driven design of \emph{CLF and stability} controller for \emph{ct-NPS}}\label{alg:ct-stab}
	\begin{algorithmic}[1]
		\Require Collected trajectories $\mathcal{U}_0^c,\mathcal{X}_0^c,\mathcal{X}_1^c$, a choice of monomials $\mathcal{M}(x)$
		\State Check that the full row-rank condition for $\mathcal{N}_0^c$ is satisfied
		\State Solve~\eqref{eq:stability-ctNPS0} and~\eqref{eq:stability-ctNPS1} for $P$ and $H(x)$ simultaneously
		\Ensure CLF $\mathcal{V}(x) = \mathcal{M}(x)^\top [\mathcal{N}_{0}^cH(x)]^{-1}\mathcal{M}(x)$ and its corresponding stability controller $u = \mathcal{U}_{0}^cH(x)[\mathcal{N}_{0}^cH(x)]^{-1}\mathcal{M}(x)$
	\end{algorithmic}
\end{algorithm}

The pseudocode for constructing the CLF and its corresponding stability controllers for ct-NPS is outlined in Algorithm~\ref{alg:ct-stab}.

\noindent {\bf Graphical User Interface (GUI) in \textsf{TRUST}.} To maximize accessibility and ensure a highly user-friendly experience, \textsf{TRUST} offers an intuitive, accessible and responsive GUI which utilizes web-based technologies to enable its use without the user having to download or install an application. The GUI enhances usability by abstracting the underlying technical complexities, allowing users to construct either a CLF or CBC using data through a \emph{reactive push-button} interface. Note that for more demanding use cases, or where the data is private, the tool can also be installed on local hardware via Docker. While \textsf{TRUST} provides GUIs for all four system classes, each focusing on either stability or safety properties, we only depict it for dt-NPS (see Subsection~\ref{new27}), as it offers \emph{the most comprehensive inputs} to the GUI (see Figure \ref{fig:TRUST}). Labels in this figure are referenced with $<\cdot>$.

As noted earlier, this iteration of \textsf{TRUST} relies on external packages which require a license for their use of MOSEK. Users must upload their license to run the tool $<1>$. Note that the license is \emph{not} stored; it is only used on a per-session basis, meaning users must re-upload their license for each new session.  Moreover, future iterations of the tool are anticipated to include optimal implementations of other solvers; \href{CVXOPT}{https://cvxopt.org/} is already supported by our tool, but given the recognition and performance of MOSEK, as well as the free academic license available, we have focused on optimizing the tool to utilize MOSEK. As future iterations are released, the GUI will include the additional input option to select a preferred solver.

 \textsf{TRUST} supports both discrete-time and continuous-time system classes $<2>$ as well as linear and nonlinear polynomial models $<3>$. It can design controllers to enforce either stability or safety properties $<4>$. Users may upload datasets for $\mathcal{X}_0, \mathcal{U}_0,$ and $\mathcal{X}_1$or enter them manually $<5>$. A link to the tool's source code is provided in the footer $<6>$. The GUI is dynamically rendered, hiding unnecessary elements to avoid user confusion.

The system dimension is automatically detected based on the shape of $\mathcal{X}_0$ $<7>$. For nonlinear polynomial systems, users must specify the monomials $\mathcal M(x)$ $<8>$, separated by semicolons. 
Note that SymPy notation requires ``$*$'' for \emph{multiplication} and ``$**$'' for \emph{exponents}. For dt-NPS, users also need to define their desired $ \Theta(x)$ (cf.~\eqref{eq:nonlinear-M=thetax}) in $<9>$.  If desired, users can enable an ``autofill'' option for $ \Theta(x)$ in $<9>$, allowing \textsf{TRUST} to automatically compute its value (cf. Remark~\ref{Re_Theta}).
If the property is safety, $\textsf{TRUST}$ uses hypercubes to define the state space $<10>$, initial sets $<11>$, and one or more unsafe sets $<12>$, with both lower and upper bounds required for each dimension. For stability properties, items $<10> - <12>$ are hidden.

Finally, the output is displayed in the far-right section, showing a brief summary with the total computation time and peak memory usage $<13>$. If \textsf{TRUST} fails to find a valid solution, this is prominently displayed in red in the ``INFO'' section, and no other output is shown. For successful runs, the CLF or CBC is designed, along with the underlying matrix $P$ $<14>$. This is followed by the controller, including the matrix $H$ $<15>$. For the CBC, the designed level sets are also displayed $<16>$. All results are formatted in Python syntax.

To run the tool, the user clicks the ``Calculate'' button to initiate computation or ``Reset'' to clear their input $<17>$. The keyboard shortcut \texttt{Cmd+Enter} or \texttt{Ctrl+Enter}, depending on the operating system, can also start the computation.

\noindent {\bf Folder Structure in \textsf{TRUST}.}  Built with industry-standard software engineering practices, \textsf{TRUST} follows the typical Model-View-Controller folder structure. The root directory contains essential files such as the project \texttt{LICENSE}, a \texttt{README.md}, and configuration documents including \texttt{docker-compose.yml}, \texttt{requirements.txt}, and the \texttt{.env} setup file. The core application logic is organized within the \textsf{app} folder, housing controller classes in \textsf{http/controllers/} and data models in \textsf{models/}. For the frontend GUI, the \textsf{vite} directory includes the HTML, CSS, and VueJS files, along with configuration settings for building UI components. While the tool does not use a database, storage needs are handled by the \textsf{storage} directory. Unit tests are structured under \texttt{tests/}, ensuring comprehensive coverage across components. Docker configurations and startup scripts are located in the \textsf{docker} folder, enabling straightforward deployment via Dockerfiles and Compose files. This structure not only enhances maintainability but also leverages modern development practices to streamline integration and testing processes.

\begin{lstlisting}[
	language=bash,label=lst:folder_structure,caption=Folder Structure in \textsf{TRUST}.]
	TRUST/
	- LICENSE
	- README.md
	- app/
	- http/
	- controllers/
	- dashboard_controller.py
	- models/
	- __init__.py
	- barrier.py
	- safety_barrier.py
	- stability.py
	- docker/
	- flask
	- Dockerfile
	- start-container
	- supervisord.conf
	- docker-compose.yml
	- main.py
	- node_modules/
	- requirements.txt
	- storage/
	- tests/
	- __init__.py
	- http/
	- models/
	- pytest.ini
	- vite/
	- css/
	- index.html
	- js/
	- main.js
	- node_modules/
	- package-lock.json
	- package.json
	- postcss.config.js
	- tailwind.config.js
	- vite.config.js
\end{lstlisting}

\noindent {\bf Error Handling.}   \textsf{TRUST} is developed as a responsive and reactive Python Flask web application, offering an \emph{intuitive, user-friendly} interface that allows seamless interaction. If a user error occurs, \textsf{TRUST} provides responsive error messages to guide the user in correcting their input. Listed below are some common errors that may be returned to the user: 
\begin{enumerate}[(i)]
	\item For an invalid ``state space'', ``initial set'' or ``unsafe set(s)'': \emph{``Provided spaces are not valid. Please provide valid lower and upper bounds''.}
	\item For an invalid shape of $\Theta(x)$: \emph{``Theta\_x should be of shape ({N}, {n})''.}
	\item If monomials are provided with commas:  \emph{``Monomial terms should be split by semicolon''}; if they are not suitable for the set dimensions:  \emph{``Monomials must be in terms of x1 (to xn)''}; if some unspecified error has occurred with the monomials:  \emph{``Invalid monomial terms''}.
	\item If the rank condition is not met for nonlinear polynomial systems: \emph{``The number of samples, T, must be greater than the number of monomial terms, N''}, or \emph{``The N0 data is not full row-rank''}, depending on which part of the rank condition failed. Similarly for linear systems: \emph{``The number of samples, T, must be greater than the number of states, n''}, or \emph{``The X0 data is not full row-rank''}, again depending on which part of the rank condition failed.
	\item If data files are uploaded with an invalid format: \emph{``Unable to parse uploaded file(s)''}.
	\item If the MOSEK solver cannot find a solution for the given values: \emph{``Solution Failure''}, with a dynamic error description provided by the solver. If the MOSEK solver did find a solution but the solution does not contain an SOS decomposition: \emph{``No SOS decomposition found''} with a dynamic error description. Similarly, if the solution does not contain valid SOS constraints: \emph{``Constraints are not sum-of-squares''}.
	\item Any other errors in the tool will be caught with the generic error message: \emph{``An unknown error occurred''} and a brief description that can be reported.
\end{enumerate}

\subsection{Safety and Stability of ct-LS}\label{subsec:continuous-time-linear-systems}

We consider continuous-time linear systems (ct-LS) defined as
\begin{equation}
\label{eq:ct-LS}
\Sigma^c\!: \dot{x} = Ax+Bu
\end{equation}
where $A\in\mathbb{R}^{n\times n}$ and $B\in\mathbb{R}^{n\times m}$ are both \emph{unknown}, and $x \in X$ and $u \in U$ represent the system's state and control input, respectively, with $X \subset \mathbb{R}^n$ and $U \subset \mathbb{R}^m$ being the state and input sets.

The following lemma, borrowed from~\cite{de2019formulas}, provides a data-based representation of closed-loop ct-LS with a controller $u=Kx$, where $K\in\mathbb{R}^{m\times n}$.

\begin{lemma}[{\bf Data-based Representation of ct-LS~\cite{de2019formulas}}]\label{lem:Q-ctLS}
Let $Q$ be a $(T\times n)$ matrix such that
\begin{equation}\label{eq:Q-matrix-ctLS}
\identity_n = \mathcal{X}_{0}^cQ,
\end{equation}
with $\mathcal{X}_{0}^c$ being a full row-rank matrix.
If one synthesizes $u=Kx = \mathcal{U}_{0}^cQx$, then the closed-loop system $\dot{x}=Ax+Bu$ has the following data-based representation:
\begin{equation}
\label{eq:A+BK=X1TQ-ctLS}
\dot{x} = \mathcal{X}_{1}^cQx,\quad\text{equivalently,}\quad A+BK = \mathcal{X}_{1}^cQ.
\end{equation}
\end{lemma}

The following theorem, adapted from~\cite{wang2024convex},
  utilizes the data-driven representation of ct-LS from Lemma~\ref{lem:Q-ctLS} to design a CBC $\mathcal B(x) = x^\top P x$, with $P \succ 0$,  and a safety controller $u= \mathcal{U}_{0}^cQx$ based on the collected data.

\begin{theorem}[{\bf Data-Driven CBC for ct-LS~\cite{wang2024convex}}]
\label{thm:data-ctLS-CBC}
Consider the ct-LS in \eqref{eq:ct-LS} with unknown matrices $A$, $B$, and its data-based representation $\dot{x}=\mathcal{X}_{1}^cQx$.
Let $X_I, X_O \subset X$ represent the initial and unsafe regions of the ct-NPS, respectively. Suppose there exist constants $\gamma,\lambda \in \mathbb R^+$, with $\lambda > \gamma$, and a  matrix $H\in\mathbb{R}^{T\times n}$ such that the following constraints are fulfilled:
\begin{subequations}
\begin{align}
\label{eq:ct_CBC_data0}
\mathcal{X}_{0}^cH=P^{-1}&, \quad\quad\quad\quad\text{with} ~P \succ 0,\\
\label{eq:ct_CBC_data1}
x^\top [\mathcal{X}_{0}^cH]^{-1}x &\leq \gamma, \quad\quad\quad\forall x\in X_I, \\
\label{eq:ct_CBC_data2}
x^\top [\mathcal{X}_{0}^cH]^{-1}x &\geq \lambda, \quad\quad\quad\forall x\in X_O,\\
\label{eq:ct_CBC_data3}
-[\mathcal{X}_{1}^cH + H^\top\mathcal{X}_{1}^{c\top}] &\geq 0.
\end{align}
\end{subequations}
Then $\mathcal{B}(x) = x^\top [\mathcal{X}_{0}^cH]^{-1}x$ is a CBC and $u = \mathcal{U}_{0}^cH[\mathcal{X}_{0}^cH]^{-1}x$ is its corresponding safety controller for the unknown ct-LS.
\end{theorem}

Since condition \eqref{eq:ct_CBC_data3} for the ct-LS reduces to a linear matrix inequality (LMI), it can be solved using semidefinite programming (SDP) solvers such as \textsf{SeDuMi}~\citep{sturm1999using}. Algorithm~\ref{alg:ct-linear} details the steps required to design a CBC and its safety controller based entirely on data for ct-LS.

The following theorem, borrowed from~\cite{de2019formulas},
 provides the required conditions for designing a CLF $\mathcal{V}(x)=x^\top Px$ with $P \succ 0$, along with a stability controller $u= \mathcal{U}_{0}^cQx$ based on collected data.
 
 \begin{algorithm}[t!]
 	\caption{Data-driven design of \emph{CBC and safety} controller for \emph{ct-LS}}\label{alg:ct-linear}
 	\begin{algorithmic}[1]
 		\Require Regions of interest $X,X_I,X_O$, collected trajectories $\mathcal{U}_0^c,\mathcal{X}_0^c,\mathcal{X}_1^c$
 		\State Check that the full row-rank condition for $\mathcal{X}_0^c$ is satisfied
 		\State Solve~\eqref{eq:ct_CBC_data0}  and~\eqref{eq:ct_CBC_data3} for $P$ and $H$, simultaneously, using SDP solvers
 		\State Given the constructed $H$, solve \eqref{eq:ct_CBC_data1} and \eqref{eq:ct_CBC_data2} via SOS optimization to design $\gamma$ and $\lambda$, where $\lambda >\gamma$
 		\Ensure CBC $\mathcal{B}(x) = x^\top [\mathcal{X}_{0}^cH]^{-1}x$ and its corresponding safety controller $u = \mathcal{U}_{0}^cH[\mathcal{X}_{0}^cH]^{-1}x$
 	\end{algorithmic}
 \end{algorithm}

\begin{theorem}[{\bf Data-Driven CLF for ct-LS~\cite{de2019formulas} }]\label{thm:ctLS-stable}
	Consider the ct-LS in \eqref{eq:ct-LS} with unknown matrices $A$, $B$, and its data-based representation $\dot{x}=\mathcal{X}_{1}^cQx$. Suppose there exists a matrix $H\in\mathbb{R}^{T\times n}$ such that the following constrains are satisfied:
	\begin{subequations}
		\begin{align}
			\label{eq:stability-ctLS0}
			\mathcal{X}_{0}^cH=P^{-1}&, \quad\quad\text{with} ~P \succ 0,\
			\\\label{eq:stability-ctLS1}
			\mathcal{X}_{1}^cH + H^\top\mathcal{X}_{1}^{c\top}&\prec 0.
		\end{align}
	\end{subequations}
Then $\mathcal{V}(x) = x^\top [\mathcal{X}_{0}^cH]^{-1}x$ is a CLF and $u = \mathcal{U}_{0}^cH[\mathcal{X}_{0}^cH]^{-1}x$ is its corresponding stability controller for the unknown ct-LS.
\end{theorem}

The pseudocode for designing a CLF and its stability controller for ct-LS is provided in Algorithm~\ref{alg:ctLS-stab}.

{\bf \textsf{TRUST}  Implementation for ct-LS.}
For the ct-LS, inputs <8> and <9> in Figure~\ref{fig:TRUST} are hidden, as they are not required.

\section{Data-Driven Results for Discrete-Time Systems}

We now present the corresponding results for the data-driven safety and stability controller synthesis of \emph{discrete-time systems} with both nonlinear polynomial and linear dynamics. We remind the reader that the collected data used in this section follows the form described in~\eqref{eq:data-discrete}.

\subsection{Safety and Stability of dt-NPS}\label{new27}

\begin{definition}[{\bf dt-NPS}]
\label{def:system-description-dt-NPS}
A discrete-time nonlinear polynomial system (dt-NPS) is described by
\begin{equation}
\label{eq:dt-NPS}
\Sigma^d\!: x^+ = A\mathcal{M}(x)+Bu,
\end{equation}
where $A\in\mathbb{R}^{n\times N}$ and $B\in\mathbb{R}^{n\times m}$ are both \emph{unknown}, $\mathcal{M}(x)\in\mathbb{R}^N$ is a vector of monomials in state $x\in X$, and $u\in U$ is a control input, with $X\subset\mathbb{R}^n$ and $U\subset\mathbb{R}^m$ being the state and input sets, respectively. Note that $x^+$ denotes the state one step ahead, \emph{i.e.,} $x(k+1)$, where $k \in \mathbb{N}$.
\end{definition}

We now present the following lemma, borrowed from~\cite[Lemma 3.2]{samari2024singletrajectory}, to obtain a data-based representation of closed-loop dt-NPS in~\eqref{eq:dt-NPS} with a controller $u=K(x)x$, where $K(x)\in\mathbb{R}^{m\times n}$ is a matrix polynomial.

\begin{lemma}[{\bf Data-based Representation of dt-NPS~\cite{samari2024singletrajectory}}]
\label{lem:Q-dtNPS}
Let $Q(x)$ be a $(T\times n)$ matrix polynomial such that
\begin{equation}
\label{eq:nonlinear-theta=NQ}
\Theta(x) = \mathcal{N}_{0}^dQ(x),
\end{equation}
where $\Theta(x)$ is an $(N\times n)$ matrix polynomial such that
\begin{equation}\label{eq:nonlinear-M=thetax}
\mathcal{M}(x)=\Theta(x)x,
\end{equation}
and $\mathcal{N}^d_{0}$ is an $(N\times T)$ full row-rank matrix as in \eqref{New3}.
If one synthesizes $u=K(x)x = \mathcal{U}_{0}^dQ(x)x$, then the closed-loop system $x^+=A\mathcal{M}(x)+Bu$ has the following data-based representation:
\begin{equation}\label{eq:nonlinearA+BK=X1TQ}
x^+ = \mathcal{X}_{1}^dQ(x)x, \quad\text{equivalently}\quad A\Theta(x)+BK(x) = \mathcal{X}_{1}^dQ(x).
\end{equation}
\end{lemma}

Using the above lemma, we now have a data-driven representation of dt-NPS in the form $\mathcal{X}_{1}^dQ(x)x$, which is utilized to design a CBC and its safety controller, as outlined in the following theorems~\cite[Theorem 3.5]{samari2024singletrajectory}.

\begin{algorithm}[t!]
	\caption{Data-driven design of \emph{CLF and stability} controller for \emph{ct-LS}}\label{alg:ctLS-stab}
	\begin{algorithmic}[1]
		\Require Collected trajectories $\mathcal{U}_0^c,\mathcal{X}_0^c,\mathcal{X}_1^c$
		\State Check that the full row-rank condition for $\mathcal{X}_0^c$ is satisfied
		\State Solve~\eqref{eq:stability-ctLS0}  and~\eqref{eq:stability-ctLS1} for $P$ and $H$, simultaneously, using SDP solvers
		\Ensure CLF $\mathcal{V}(x) = x^\top [\mathcal{X}_{0}^cH]^{-1}x$ and its corresponding stability controller $u = \mathcal{U}_{0}^cH[\mathcal{X}_{0}^cH]^{-1}x$
	\end{algorithmic}
\end{algorithm}

\begin{theorem}[{\bf Data-Driven CBC for dt-NPS~\cite{samari2024singletrajectory}}]
\label{thm:data-dtNPS-CBC}
Consider the dt-NPS in~\eqref{eq:dt-NPS} with unknown matrices $A$, $B$, and its data-based representation $x^+=\mathcal{X}_{1}^dQ(x)x$.
Let $X_I, X_O \subset X$ represent the initial and unsafe regions of the dt-NPS, respectively. Suppose there exist constants $\gamma,\lambda \in \mathbb R^+$, with $\lambda > \gamma$, and a  matrix polynomial $H(x)\in\mathbb{R}^{T\times n}$ such that the following constraints are fulfilled:
\begin{subequations}
\begin{align}
\label{eq:nonlinear-safe_CBC_data0}
	\mathcal{N}_{0}^dH(x)= \Theta(x)P^{-1} &, \quad\quad\quad\text{with} ~P \succ 0,\\
\label{eq:nonlinear-safe_CBC_data1}
x^\top  [\Theta^\dagger \mathcal{N}_{0}^dH(x)]^{-1}x \leq \gamma, &\quad\quad\quad\quad\forall x\in X_I, \\
\label{eq:nonlinear-safe_CBC_data2}
x^\top [\Theta^\dagger \mathcal{N}_{0}^dH(x)]^{-1}x \geq \lambda, &\quad\quad\quad\quad\forall x\in X_O,\\
\label{eq:nonlinear-safe_CBC_data3}
\begin{bmatrix}
P^{-1}              & \mathcal{X}_{1}^dH(x) \\
H(x)^\top \mathcal{X}_{1}^{d\top}  & P^{-1}
\end{bmatrix} &\geq 0, \quad\quad\forall x\in X.
\end{align}
\end{subequations}
Then $\mathcal{B}(x) = x^\top  [\Theta^\dagger \mathcal{N}_{0}^dH(x)]^{-1}x$ is a CBC and $u = \mathcal{U}_{0}^dH(x) [\Theta^\dagger \mathcal{N}_{0}^d\\H(x)]^{-1}x$ is its corresponding safety controller for the unknown dt-NPS, with $\Theta^\dagger $ being the left pseudoinverse of $\Theta$.
\end{theorem}

The SOS optimization program can be applied to enforce conditions \eqref{eq:nonlinear-safe_CBC_data0}-\eqref{eq:nonlinear-safe_CBC_data3}, similar to those in Lemma~\ref{Lemma_ct-NPS}. The corresponding pseudocode for designing a CBC and its safety controller is provided in Algorithm~\ref{alg:dt-nonlinear}.

\begin{remark}\label{Re_Theta}
	Note that the choice of $\Theta(x)$ satisfying the equality condition in  \eqref{eq:nonlinear-M=thetax} is not unique, and different choices may influence the proposed conditions in Theorem \ref{thm:data-dtNPS-CBC}. To accommodate this flexibility, we allow the user to input this $\Theta(x)$ as an additional parameter (see $<9>$ in Figure  \ref{fig:TRUST}). Alternatively, the user can enable the autofill option, prompting the tool to solve \eqref{eq:nonlinear-M=thetax} and design $\Theta(x)$ automatically.
\end{remark}

The following theorem, adapted from~\cite{samari2024singletrajectory}, provides the sufficient conditions for designing a CLF $\mathcal{V}(x)=x^\top Px$, with $P \succ 0$, along with a stability controller $u= \mathcal{U}_{0}^dQ(x)x$ based on collected data.

\begin{theorem}[{\bf Data-Driven CLF for dt-NPS \cite{samari2024singletrajectory}}]\label{thm:dtNPS-stable}
	Consider the dt-NPS in \eqref{eq:dt-NPS} with unknown matrices $A$, $B$, and its data-based representation $x^+=\mathcal{X}_{1}^dQ(x)x$. Suppose there exists a polynomial matrix $H(x)\in\mathbb{R}^{T\times n}$ such that the following constrains are satisfied:
	\begin{subequations}
		\begin{align}
			\label{eq:schur-stability-dtNPS0}
			\mathcal{N}_{0}^dH(x)= \Theta(x)P^{-1} , &\quad\quad\text{with} ~P \succ 0,\
			\\\label{eq:schur-stability-dtNPS1}
			\begin{bmatrix}
				P^{-1}              & \mathcal{X}_{1}^dH(x) \\
				H(x)^\top \mathcal{X}_{1}^{d\top}  & P^{-1}
			\end{bmatrix} \succ 0.
		\end{align}
	\end{subequations}
Then $\mathcal{V}(x) = x^\top  [\Theta^\dagger \mathcal{N}_{0}^dH(x)]^{-1}x$ is a CLF and $u = \mathcal{U}_{0}^dH(x) [\Theta^\dagger \mathcal{N}_{0}^d\\H(x)]^{-1}x$ is its corresponding stability controller for the unknown dt-NPS, with $\Theta^\dagger $ being the left pseudoinverse of $\Theta$.
\end{theorem}

The pseudocode for constructing the CLF and its stability controllers for dt-NPS is outlined in Algorithm~\ref{alg:dt-stab}.

\begin{algorithm}[t]
	\caption{Data-driven design of \emph{CBC and safety} controller for \emph{dt-NPS}}\label{alg:dt-nonlinear}
	\begin{algorithmic}[1]
		\Require Regions of interest $X,X_I,X_O$, collected trajectories $\mathcal{U}_0^d,\mathcal{X}_0^d,\mathcal{X}_1^d$, a choice of monomials $\mathcal{M}(x)$
		\State Check that the full row-rank condition for $\mathcal{N}_0^d$ is satisfied
		\State Provide $\Theta(x)$ or select the autofill option to let \textsf{TRUST} solve it based on~\eqref{eq:nonlinear-M=thetax} 
		\State Solve~\eqref{eq:nonlinear-safe_CBC_data0} and~\eqref{eq:nonlinear-safe_CBC_data3} for $P$ and $H(x)$, simultaneously\footnotemark[3]
		\State Given the constructed $H(x)$, solve \eqref{eq:nonlinear-safe_CBC_data1} and \eqref{eq:nonlinear-safe_CBC_data2} to design $\gamma$ and $\lambda$, where $\lambda >\gamma$
		\Ensure CBC $\mathcal{B}(x) = x^\top  [\Theta^\dagger \mathcal{N}_{0}^dH(x)]^{-1}x$ and its corresponding safety controller $u = \mathcal{U}_{0}^dH(x) [\Theta^\dagger \mathcal{N}_{0}^dH(x)]^{-1}x$
	\end{algorithmic}
\end{algorithm}
\footnotetext[3]{To satisfy conditions~\eqref{eq:nonlinear-safe_CBC_data0} and~\eqref{eq:nonlinear-safe_CBC_data3}, we define $Z = P^{-1}$ and enforce that it is a \emph{symmetric positive-definite} matrix, \emph{i.e.,} $Z\succ 0$. Once conditions~~\eqref{eq:nonlinear-safe_CBC_data0},\eqref{eq:nonlinear-safe_CBC_data3} are met and $Z$ is designed, the matrix $P$ is computed as the inverse of $Z$, \emph{i.e.,} $P = Z^{-1}$.}

\subsection{Safety and Stability of dt-LS}

We consider discrete-time linear systems (dt-LS), defined as
\begin{equation}
\label{eq:dt-LS}
\Sigma^d: x^+ = Ax+Bu,
\end{equation}
where $A\in\mathbb{R}^{n\times n}$ and $B\in\mathbb{R}^{n\times m}$ are both \emph{unknown}.

Similar to Lemma~\ref{lem:Q-ctLS} under condition~\eqref{eq:Q-matrix-ctLS} and full row-rank assumption of $\mathcal{X}_{0}^d$, it can be shown that the closed-loop system $x^+=Ax+Bu$ has the following data-based representation~\cite[Theorem 2]{de2019formulas}:
\begin{equation}
\label{eq:A+BK=X1TQ-dtLS}
x^+ = \mathcal{X}_{1}^dQx,\quad\text{equivalently,}\quad A+BK = \mathcal{X}_{1}^dQ.
\end{equation}

The following theorem, adapted from~\cite[Theorem 2]{de2019formulas}, utilizes the data-driven representation of dt-LS to design a CBC $\mathcal B(x) = x^\top P x$, with $P \succ 0$,  and its safety controller $u= \mathcal{U}_{0}^dQx$ based on the collected data.

\begin{theorem}[{\bf Data-Driven CBC for dt-LS~\cite{de2019formulas}}]\label{thm:data-dtLS-CBC}
Consider the dt-LS in~\eqref{eq:dt-LS}  with unknown matrices $A$, $B$, and its data-based representation $x^+=\mathcal{X}_{1}^dQx$.
Let $X_I, X_O \subset X$ represent the initial and unsafe regions of dt-LS, respectively. Suppose there exist constants $\gamma,\lambda \in \mathbb R^+$, with $\lambda > \gamma$, and a  matrix $H\in\mathbb{R}^{T\times n}$ such that the following constraints are fulfilled:
\begin{subequations}
\begin{align}
\label{eq:safe_CBC_data0}
\mathcal{X}_{0}^dH=P^{-1}&, \quad\quad\quad\quad\text{with} ~P \succ 0,\\
\label{eq:safe_CBC_data1}
x^\top [\mathcal{X}_{0}^dH]^{-1}x \leq \gamma, &\quad\quad\quad\quad\forall x\in X_I, \\
\label{eq:safe_CBC_data2}
x^\top [\mathcal{X}_{0}^dH]^{-1}x \geq \lambda, &\quad\quad\quad\quad\forall x\in X_O,\\
\label{eq:safe_CBC_data3}
\begin{bmatrix}
P^{-1}           & \mathcal{X}_{1}^dH \\
H^\top\mathcal{X}_{1}^{d\top} & P^{-1}
\end{bmatrix} \geq 0.
\end{align}
\end{subequations}
Then $\mathcal{B}(x) = x^\top [\mathcal{X}_{0}^dH]^{-1}x$ is a CBC and $u = \mathcal{U}_{0}^dH[\mathcal{X}_{0}^dH]^{-1}x$ is its corresponding safety controller for the unknown dt-LS.
\end{theorem}

\begin{algorithm}[t]
	\caption{Data-driven design of \emph{CLF and stability} controllers for \emph{dt-NPS}}\label{alg:dt-stab}
	\begin{algorithmic}[1]
		\Require collected trajectories $\mathcal{U}_0^d,\mathcal{X}_0^d,\mathcal{X}_1^d$, a choice of monomials $\mathcal{M}(x)$
		\State Check that the full row-rank condition for $\mathcal{N}_0^d$ is satisfied
		\State Provide $\Theta(x)$ or select the autofill option to let \textsf{TRUST} solve it based on~\eqref{eq:nonlinear-M=thetax} 
		\State Solve~\eqref{eq:schur-stability-dtNPS0} and~\eqref{eq:schur-stability-dtNPS1} for $P$ and $H(x)$, simultaneously
		\Ensure CLF $\mathcal{V}(x) = x^\top  [\Theta^\dagger \mathcal{N}_{0}^dH(x)]^{-1}x$ and its corresponding stability controller $u = \mathcal{U}_{0}^dH(x) [\Theta^\dagger \mathcal{N}_{0}^dH(x)]^{-1}x$
	\end{algorithmic}
\end{algorithm}

The linear matrix (in)equalities in \eqref{eq:safe_CBC_data0},  \eqref{eq:safe_CBC_data3} can be solved using SDP solvers such as \textsf{SeDuMi}~\citep{sturm1999using}, while conditions  \eqref{eq:safe_CBC_data1},  \eqref{eq:safe_CBC_data2} can be solved using SOSTOOLS~\cite{prajna2004sostools}. The pseudocode for designing a CBC and its safety controller for dt-LS is provided in Algorithm~\ref{alg:dt-linear}.

The following theorem,  borrowed from~\cite[Theorem 2]{de2019formulas}, provides the required conditions for designing a CLF $\mathcal{V}(x)=x^\top Px$, with $P \succ 0$, along with a stability controller $u= \mathcal{U}_{0}^dQx$ based on collected data.

\begin{theorem}[{\bf Data-Driven CLF for dt-LS~\cite{de2019formulas}}]\label{thm:dtLS-stable}
	Consider the dt-LS in \eqref{eq:dt-LS} with unknown matrices $A$, $B$, and its data-based representation $x^+=\mathcal{X}_{1}^dQx$. Suppose there exists a matrix $H\in\mathbb{R}^{T\times n}$ such that the following constrains are satisfied:
	\begin{subequations}
		\begin{align}
			\label{eq:schur-stability-dtLS0}
			\mathcal{X}_{0}^dH=P^{-1}&, \quad\quad\text{with} ~P \succ 0,\
			\\\label{eq:schur-stability-dtLS1}
			\begin{bmatrix}
P^{-1}           & \mathcal{X}_{1}^dH \\
H^\top\mathcal{X}_{1}^{d\top} & P^{-1}
\end{bmatrix} \geq 0.
		\end{align}
	\end{subequations}
	Then $\mathcal{V}(x) = x^\top [\mathcal{X}_{0}^dH]^{-1}x$ is a CLF and $u = \mathcal{U}_{0}^dH[\mathcal{X}_{0}^dH]^{-1}x$ is its corresponding stability controller for the unknown dt-LS.
\end{theorem}

The pseudocode for designing a CLF and its stability controller for dt-LS is provided in Algorithm~\ref{alg:dtls-stab}.

\begin{algorithm}[t!]
	\caption{Data-driven design of \emph{CBC and safety} controller for \emph{dt-LS}}\label{alg:dt-linear}
	\begin{algorithmic}[1]
		\Require Regions of interest $X,X_I,X_O$, collected trajectories $\mathcal{U}_0^d,\mathcal{X}_0^d,\mathcal{X}_1^d$
		\State Check that the full row-rank condition for $\mathcal{X}_0^d$ is satisfied
		\State Solve~\eqref{eq:safe_CBC_data0}  and~\eqref{eq:safe_CBC_data3} for $P$ and $H$, simultaneously
		\State Given the constructed $H$, solve \eqref{eq:safe_CBC_data1} and \eqref{eq:safe_CBC_data2} via SOS optimization to design $\gamma$ and $\lambda$, where $\lambda >\gamma$
		\Ensure CBC $\mathcal{B}(x) = x^\top [\mathcal{X}_{0}^dH]^{-1}x$ and its corresponding safety controller $u = \mathcal{U}_{0}^dH[\mathcal{X}_{0}^dH]^{-1}x$
	\end{algorithmic}
\end{algorithm}

{\bf \textsf{TRUST}  Implementation for dt-LS.} 
For the dt-LS, inputs <8> and <9> in Figure~\ref{fig:TRUST} are hidden, as they are not required.

\section{Benchmarks and Evaluations}\label{sec:benchmarks-and-case-studies}

We demonstrate the effectiveness of \textsf{TRUST} through a series of physical benchmarks, covering the \emph{four classes} of dynamical systems and showcasing their respective \emph{stability or safety} properties. The mathematical models for all case studies are provided in the Appendix. However, we assume these models are unknown, relying solely on the trajectories collected from them for analysis. Table~\ref{tab:safety_stable} presents the results for the construction of CBC and safety controllers, while Table~\ref{tab:stability_stable} shows the results for the design of CLF and stability controllers. 

\begin{figure*}[h!]
	\subfloat[ct-LS: DC Motor]{\includegraphics[
		width=0.34\textwidth
		]{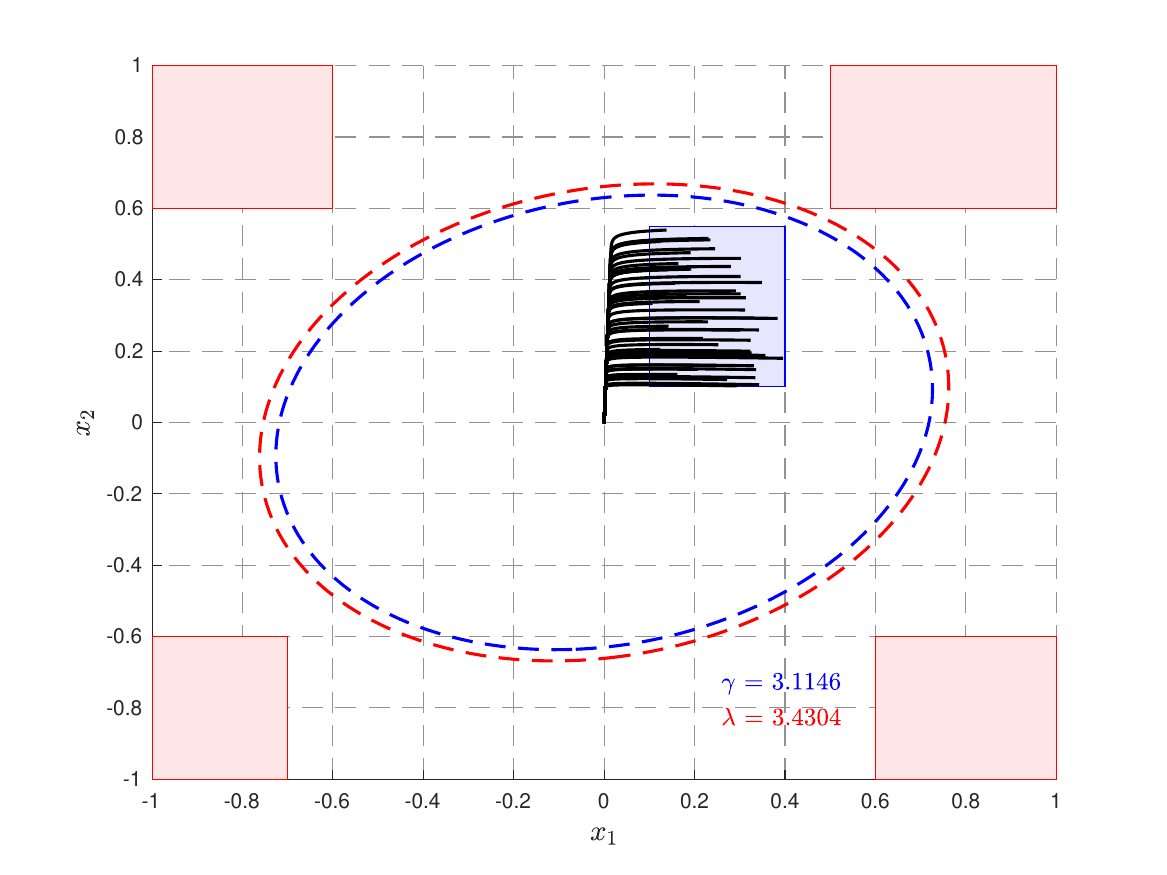}}\vspace{-0.1cm}\hspace{-0.5cm}
	\subfloat[ct-NPS: Predator-Prey]{\includegraphics[
		width=0.34\textwidth
		]{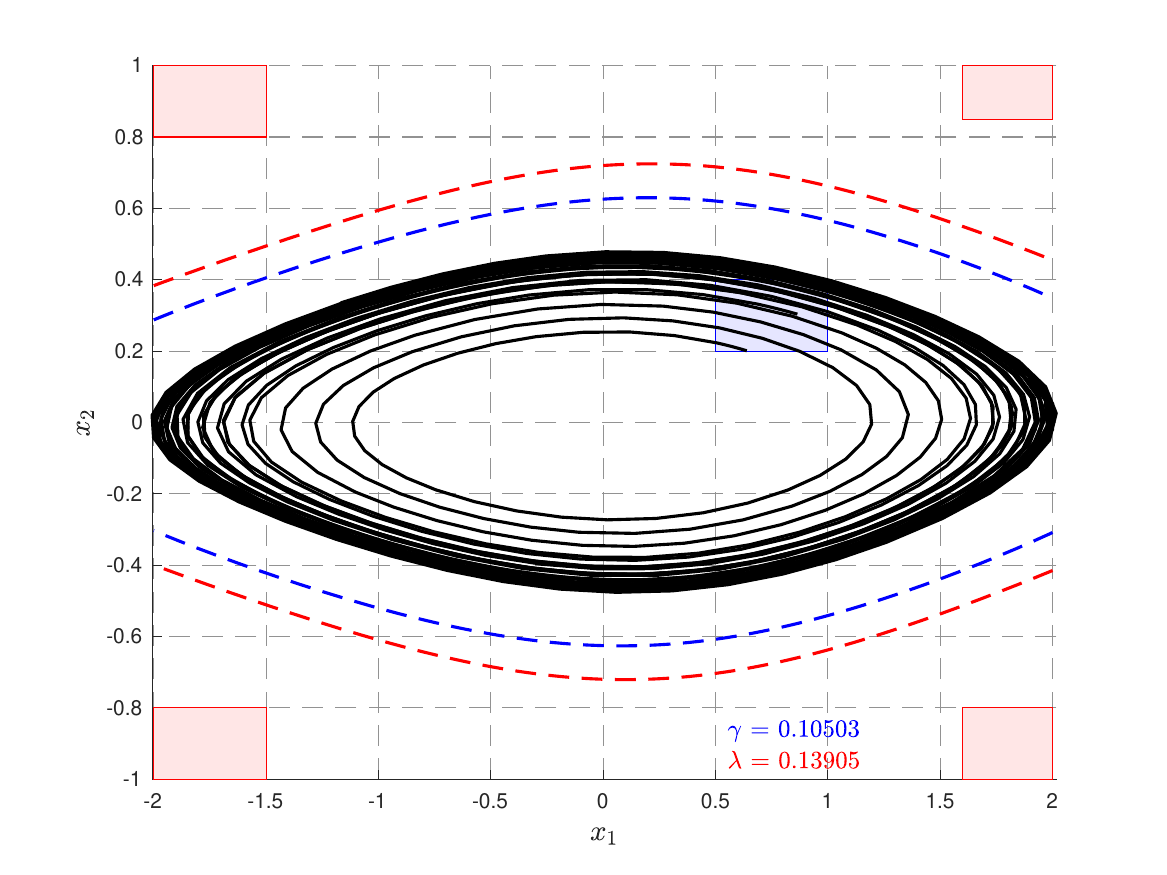}}\vspace{-0.1cm}\hspace{-0.5cm}
	\subfloat[ct-NPS: Van der Pol Oscillator]{ \includegraphics[
		width=0.34\textwidth
		]{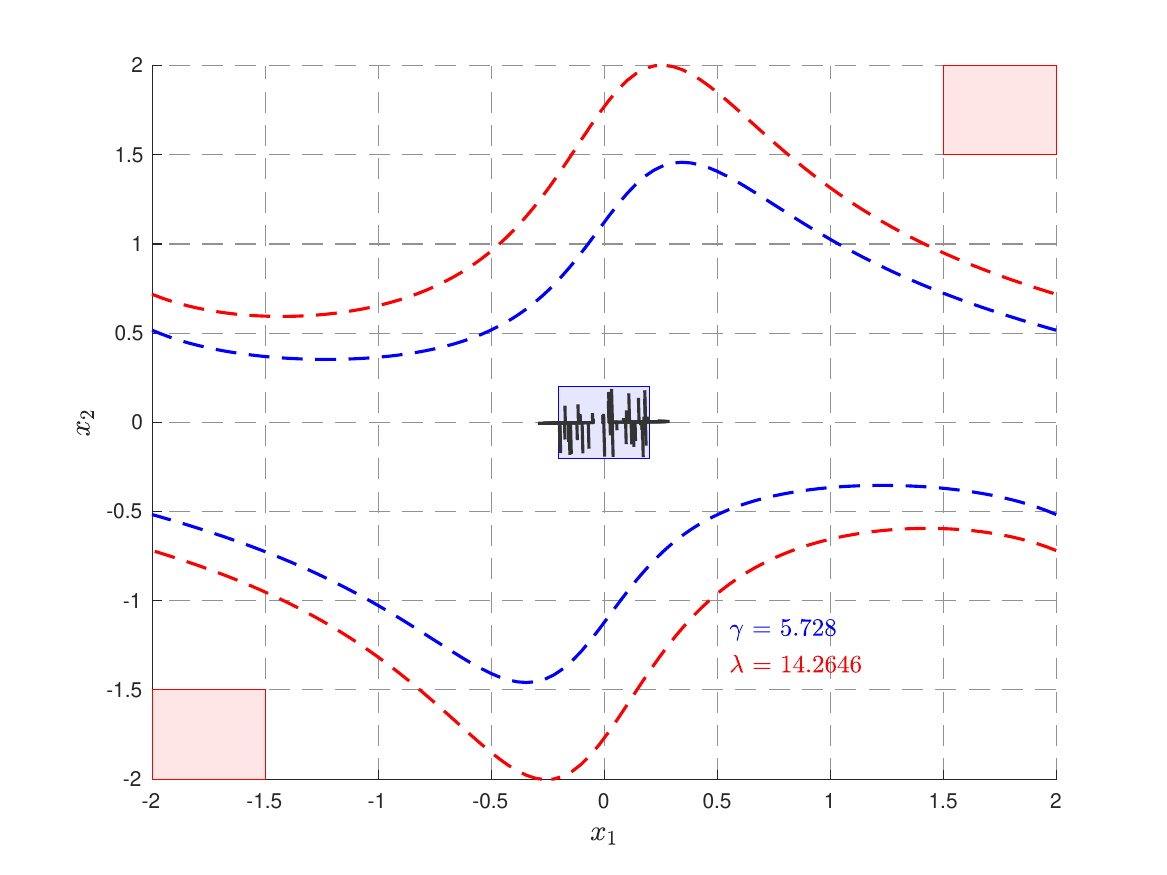}}\vspace{-0.1cm}\hspace{-0.5cm}
	\subfloat[ct-LS: Room Temperature 1]{\includegraphics[
		width=0.34\textwidth
		]{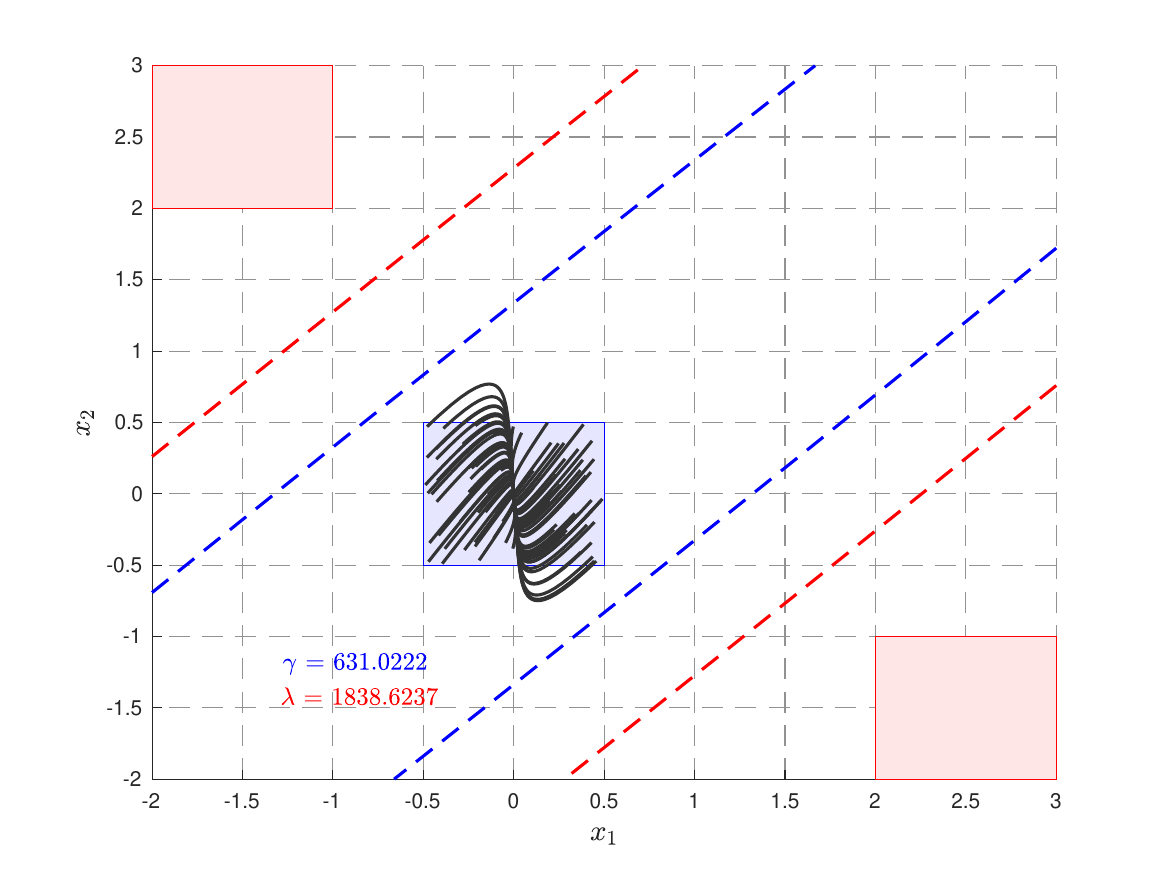}}\vspace{-0.1cm}
	\hspace{-0.5cm}
	\subfloat[ct-LS: Two Tank]{\includegraphics[
		width=0.34\textwidth
		]{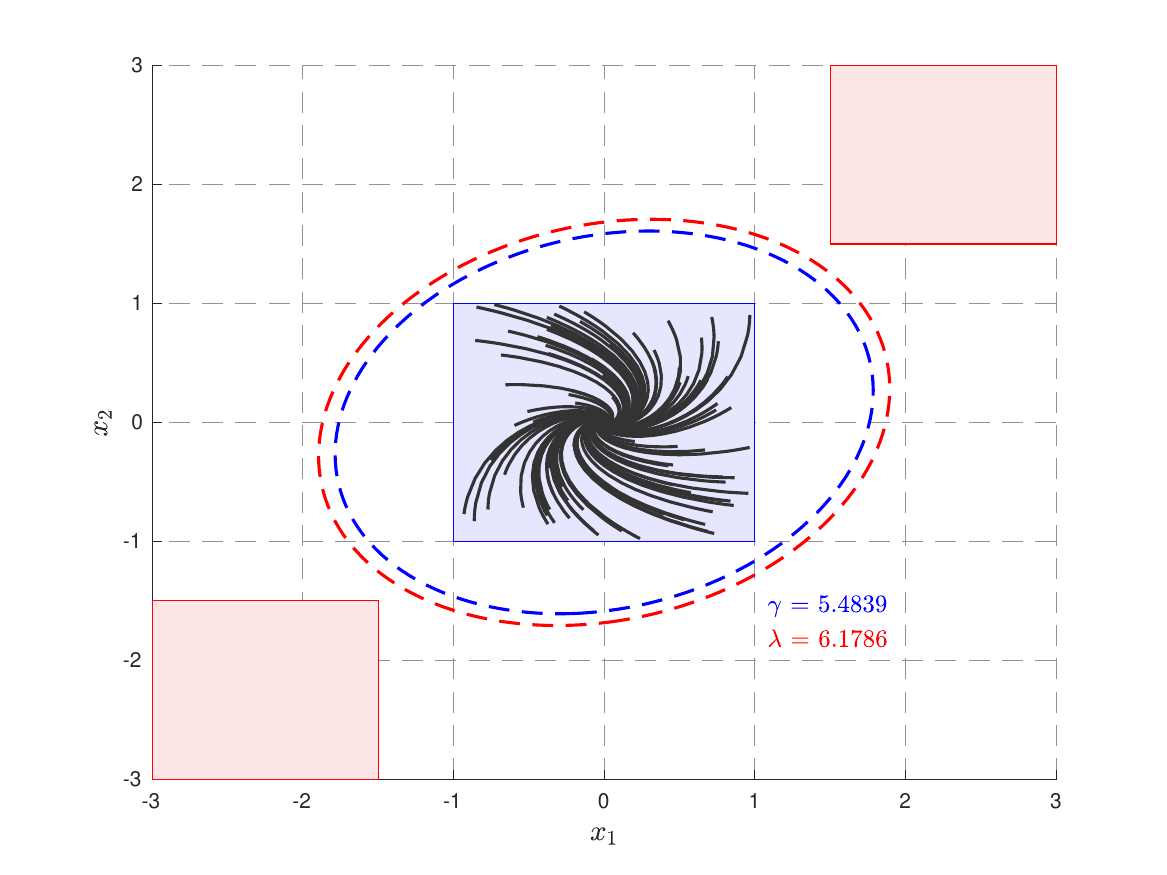}}\vspace{-0.1cm}\hspace{-0.5cm}
	\subfloat[dt-NPS: Predator-Prey]{\includegraphics[
		width=0.34\textwidth
		]{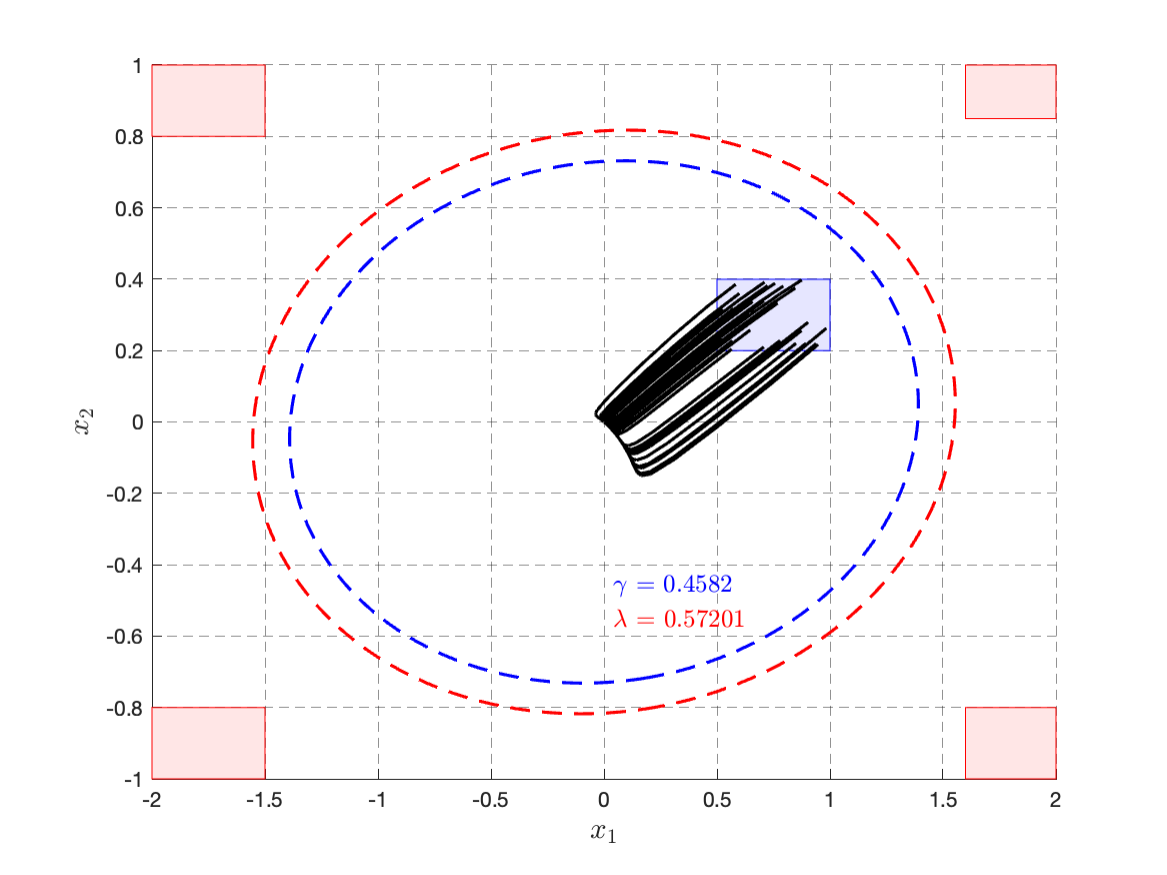}}\vspace{-0.1cm}\hspace{-0.5cm}
	\subfloat[dt-LS: DC Motor]{\includegraphics[
		width=0.34\textwidth
		]{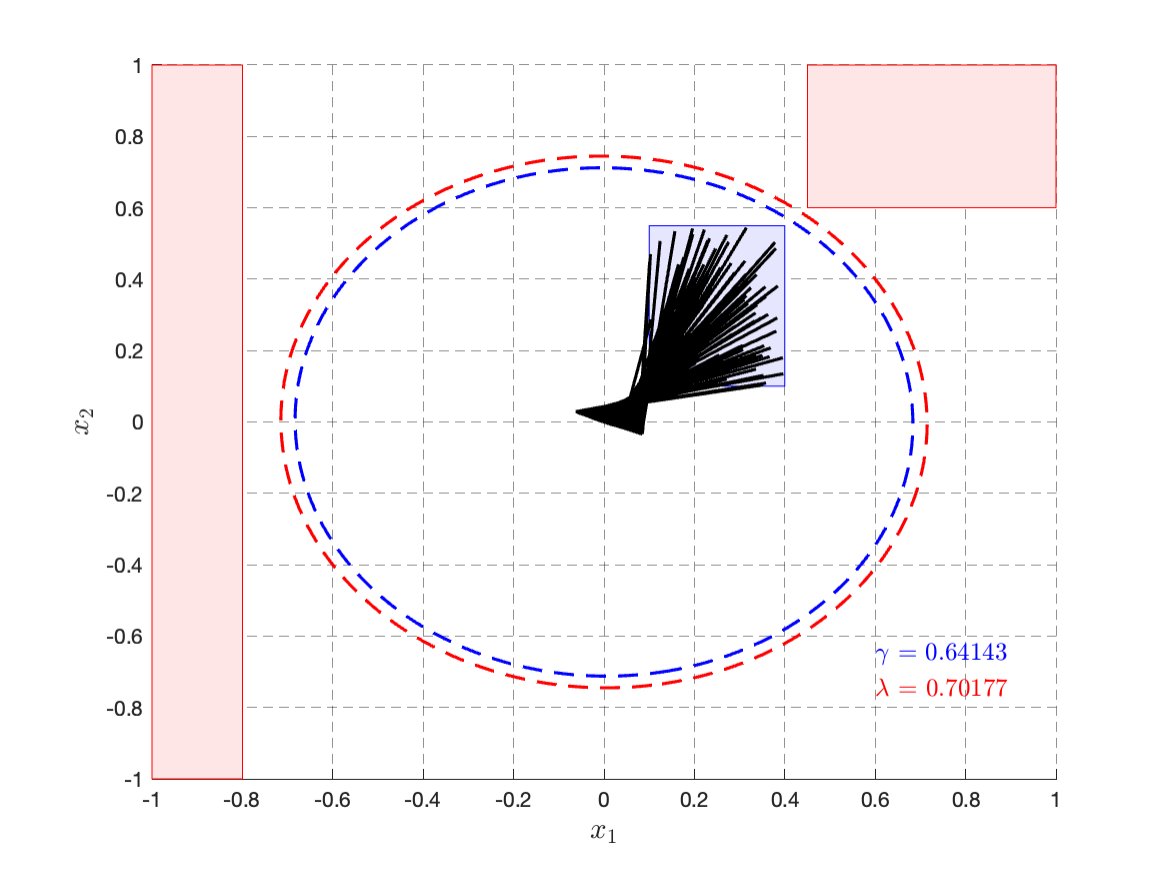}}\vspace{-0.1cm}\hspace{-0.5cm}
	\subfloat[dt-LS: Room Temperature 1]{\includegraphics[
		width=0.34\textwidth
		]{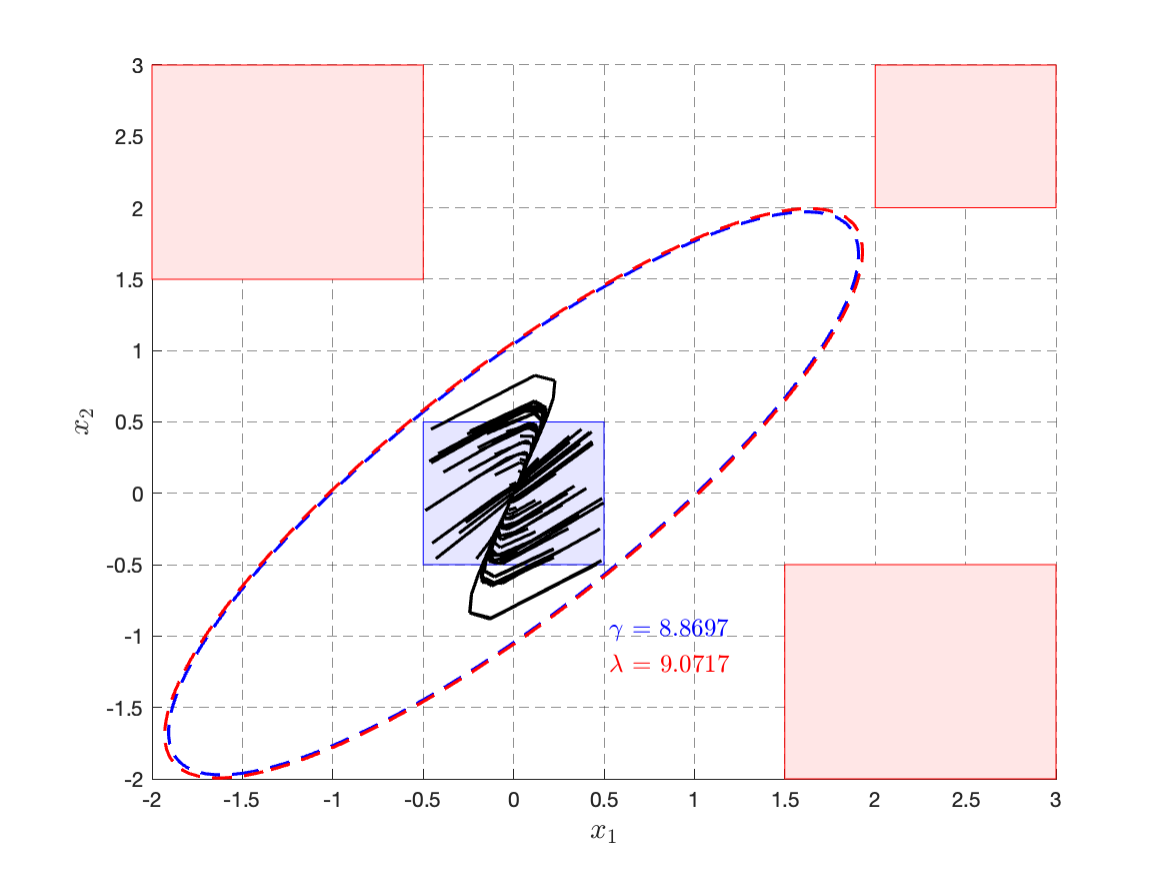}}\vspace{-0.1cm}\hspace{-0.5cm}
	\subfloat[dt-LS: Two Tank]{\includegraphics[
		width=0.34\textwidth
		]{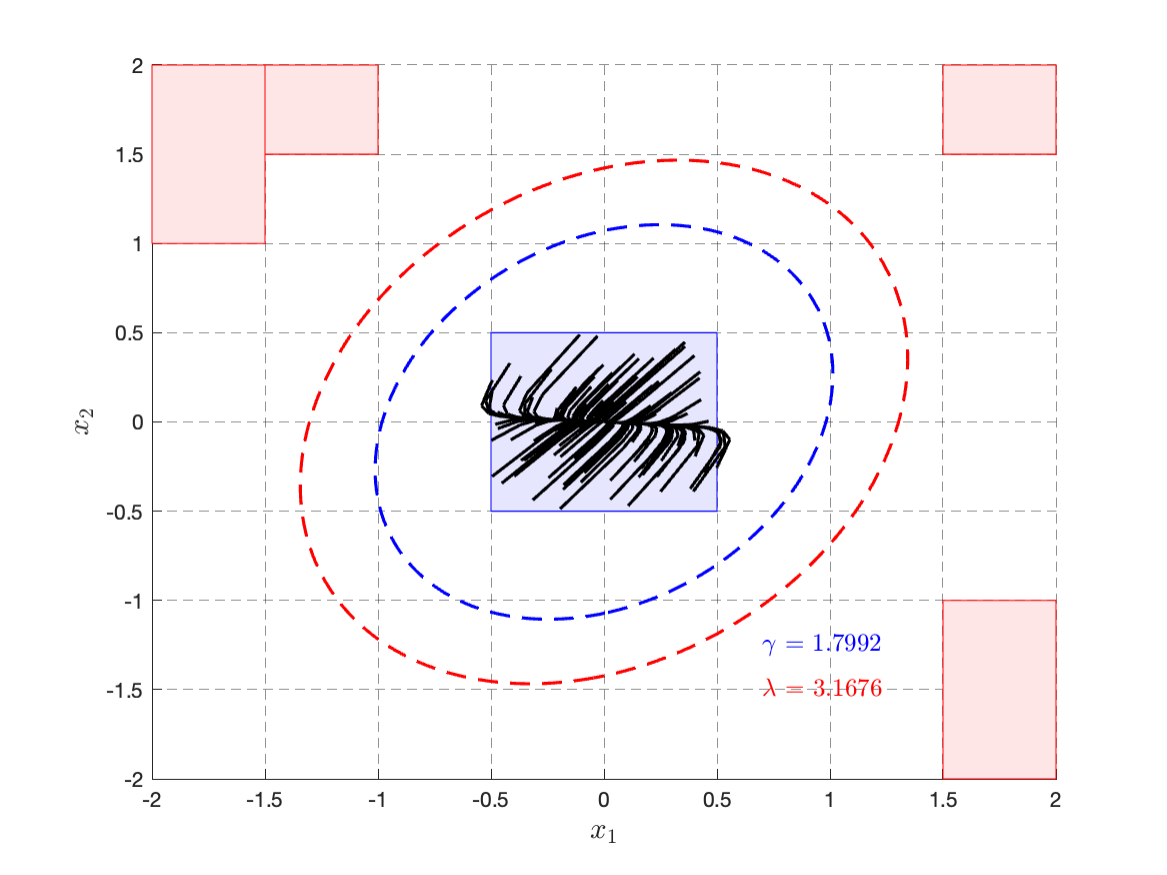}}\vspace{-0.1cm}\hspace{-0.5cm}
	\subfloat[dt-LS: Room Temperature 2]{\includegraphics[
		width=0.34\textwidth
		]{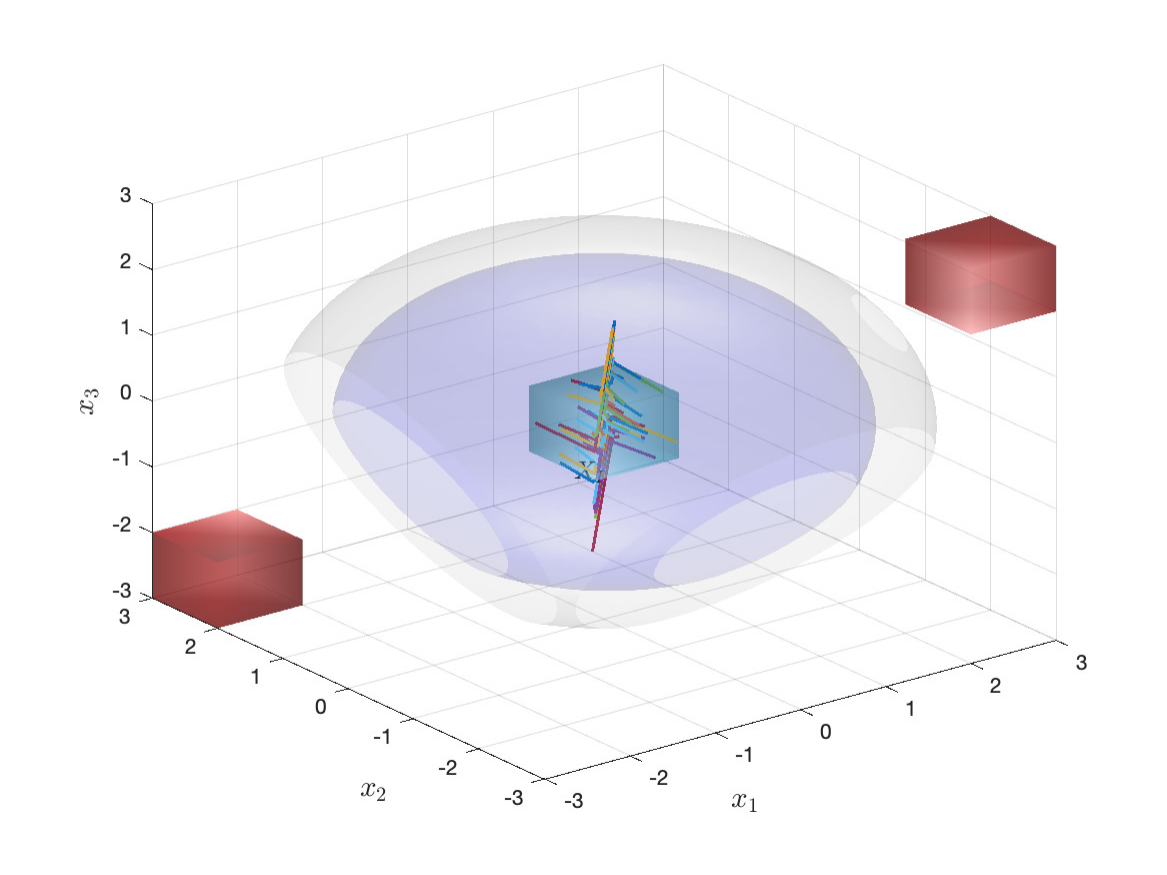}}
	\subfloat[dt-NPS: Lorenz Attractor]{\includegraphics[
		width=0.34\textwidth
		]{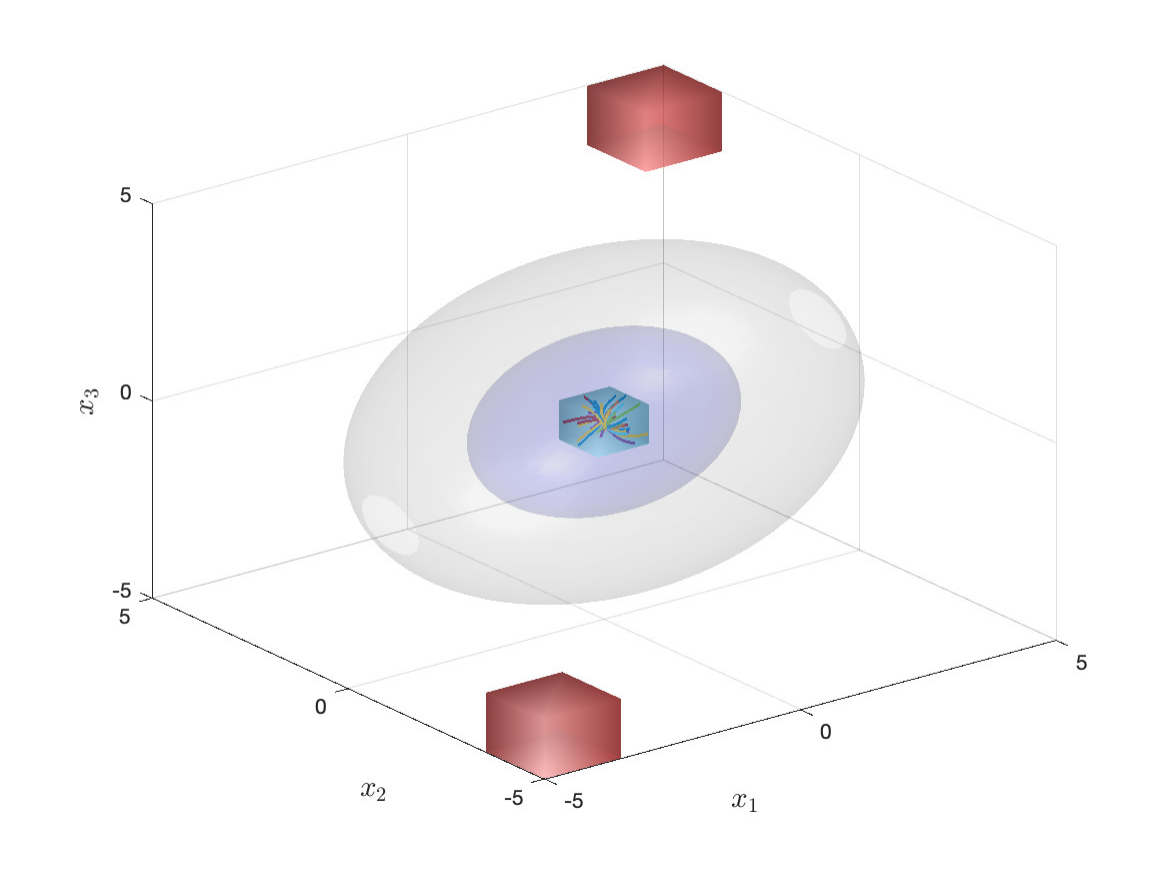}}
	\caption{\textsf{Simulation results for designing CBCs. The purple box represents the initial region, while the red boxes indicate multiple unsafe regions. For the 2D figures, the blue and red dashed lines show the initial and unsafe level sets, respectively. For the 3D figures, the purple bubble indicates the initial level set, while the gray bubble represents the unsafe level set.}}
	\label{fig:CBCs}
\end{figure*}

\begin{figure*}[t]
    \subfloat[ct-LS: DC Motor]{\includegraphics[
        width=0.34\textwidth
    ]{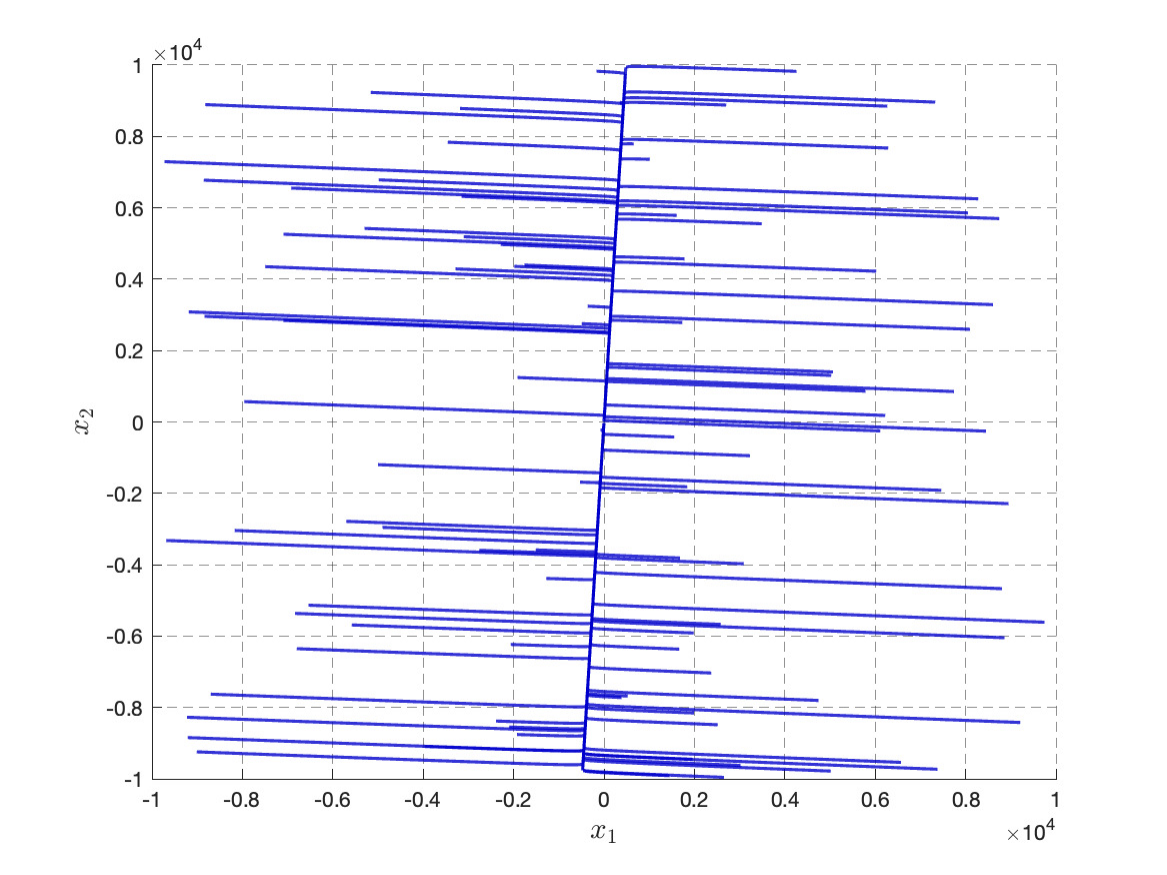}}\vspace{-0.1cm}\hspace{-0.5cm}
     \subfloat[ct-NPS: Predator Prey]{\includegraphics[
    width=0.34\textwidth
    ]{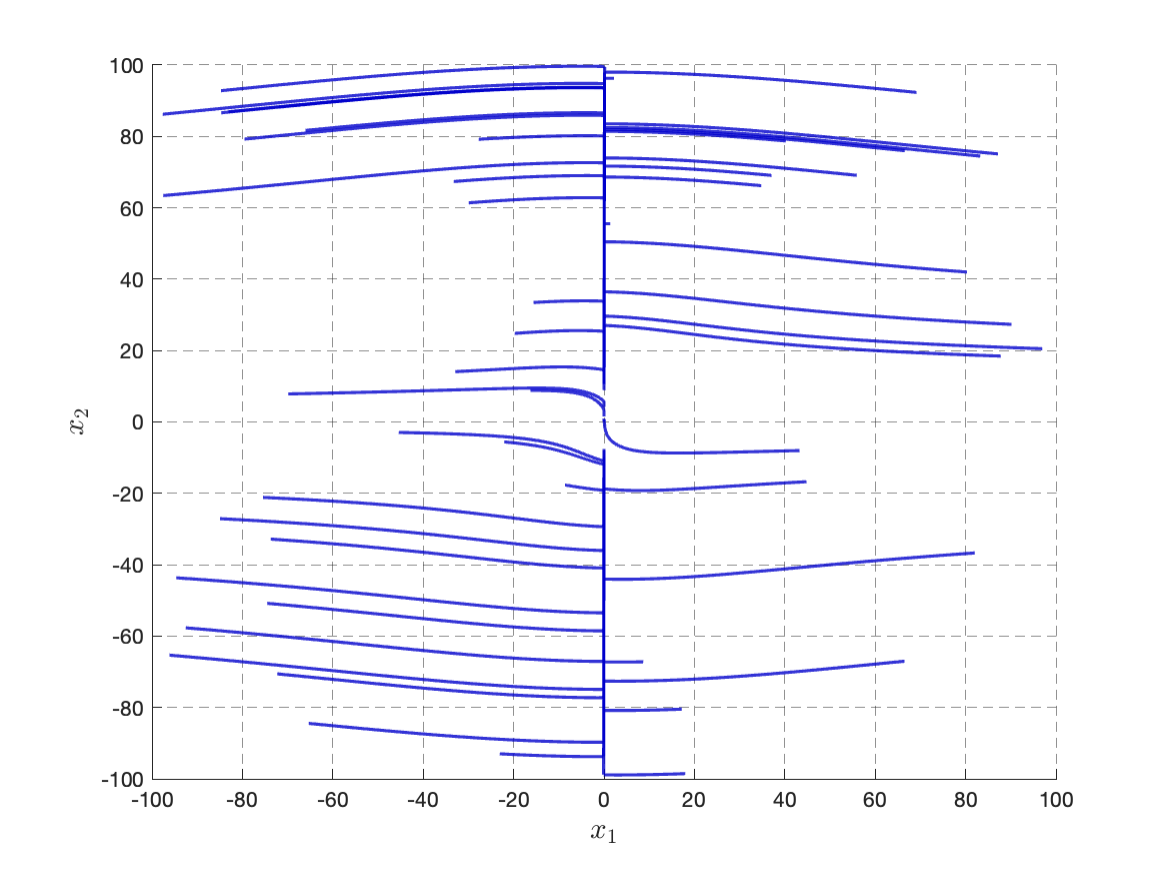}}\vspace{-0.1cm}\hspace{-0.5cm}
    \subfloat[ct-NPS: Van der Pol Oscillator]{\includegraphics[
    width=0.34\textwidth
    ]{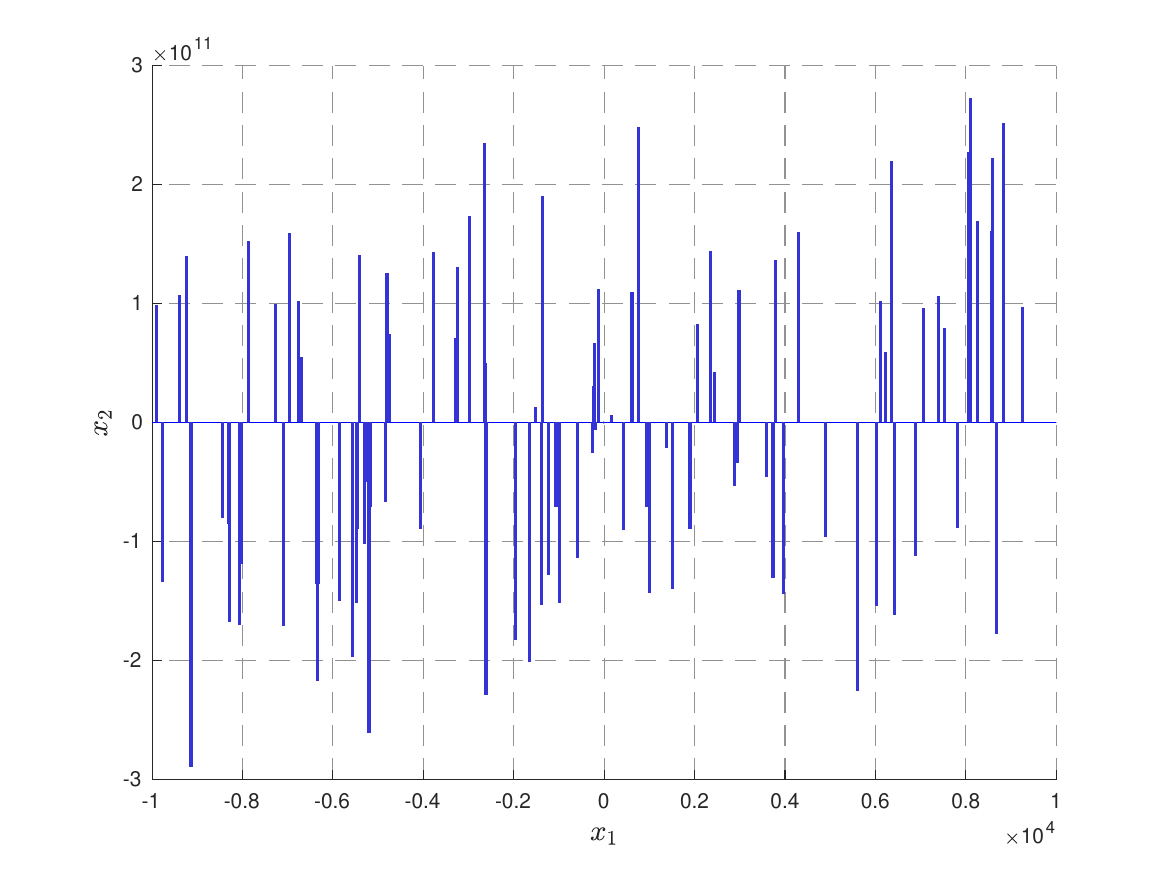}}\vspace{-0.1cm}\hspace{-0.5cm}
     \subfloat[ct-LS: Room Temperature 1]{\includegraphics[
    width=0.34\textwidth
    ]{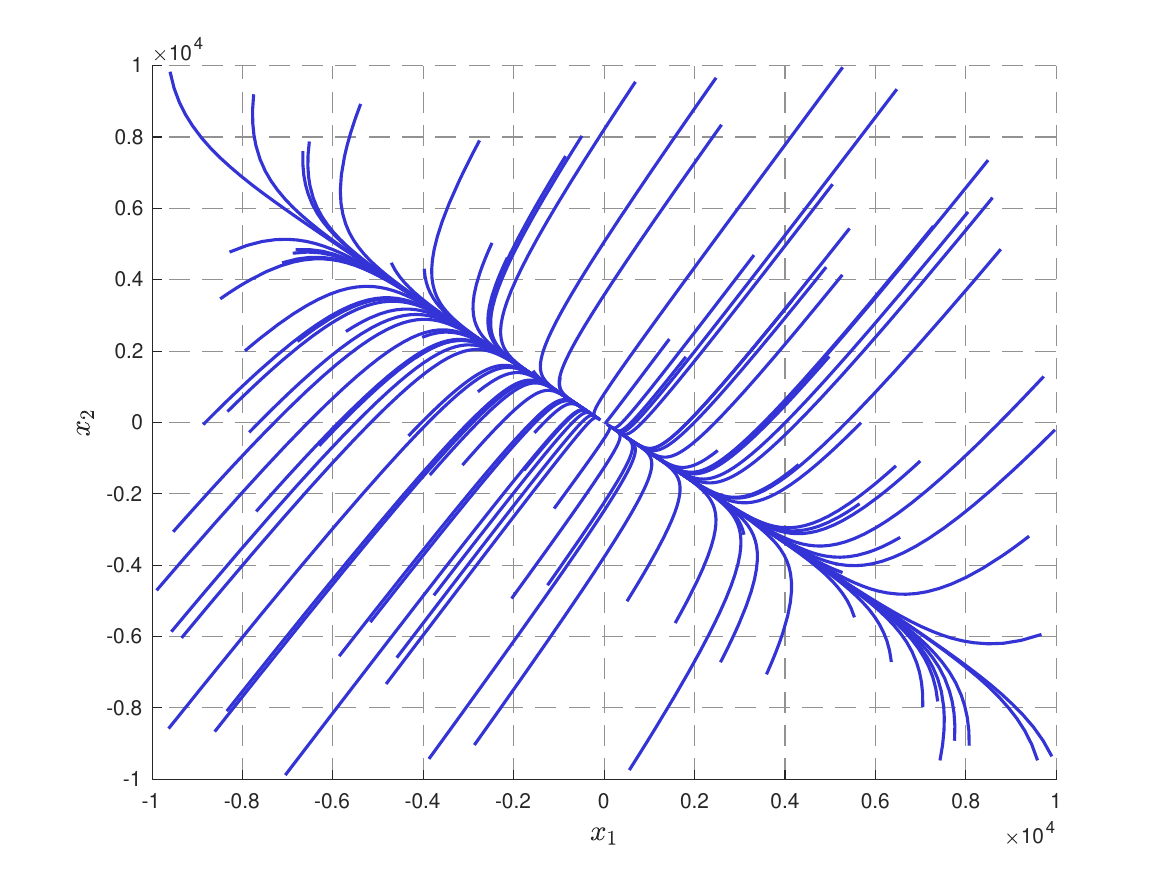}}\hspace{-0.5cm}
        \subfloat[ct-LS: Two Tank]{\includegraphics[
    width=0.34\textwidth
    ]{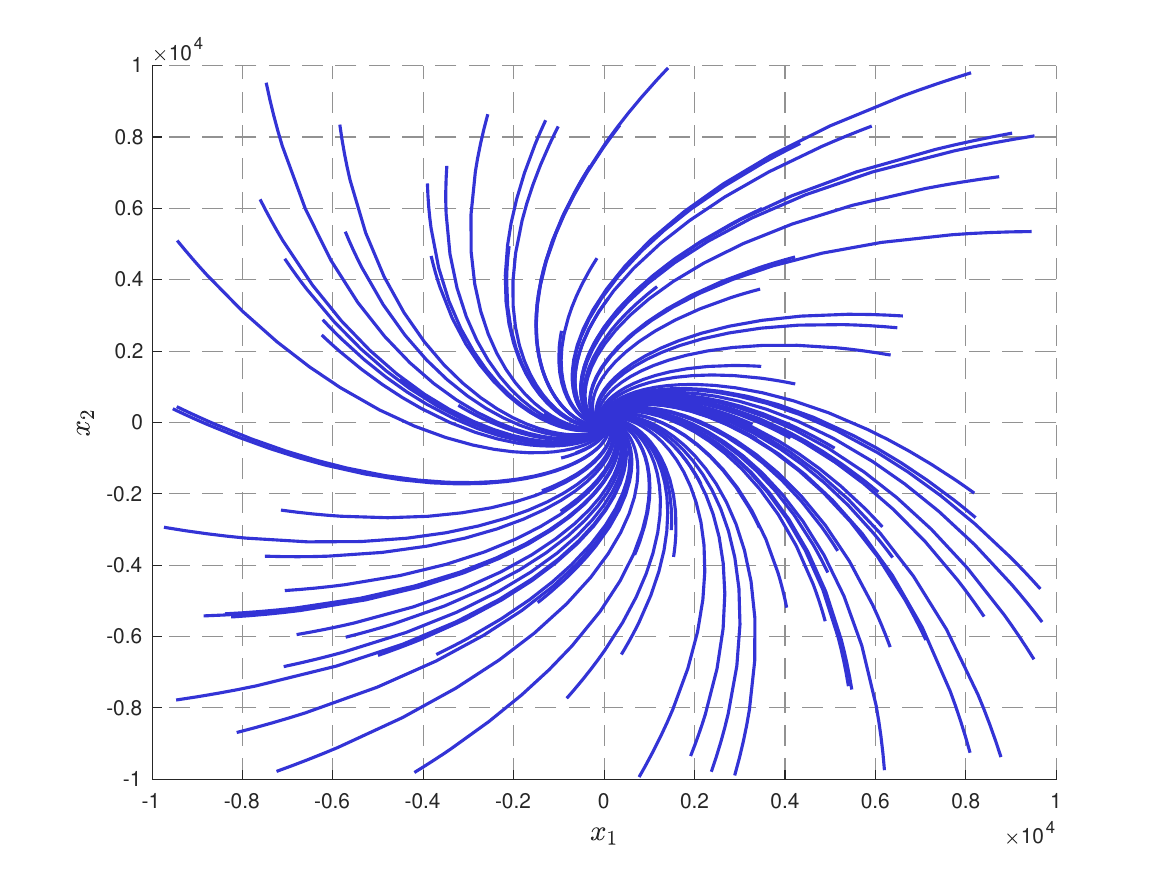}}\vspace{-0.1cm}\hspace{-0.5cm}
    \subfloat[dt-NPS: Academic System]{\includegraphics[
    width=0.34\textwidth
    ]{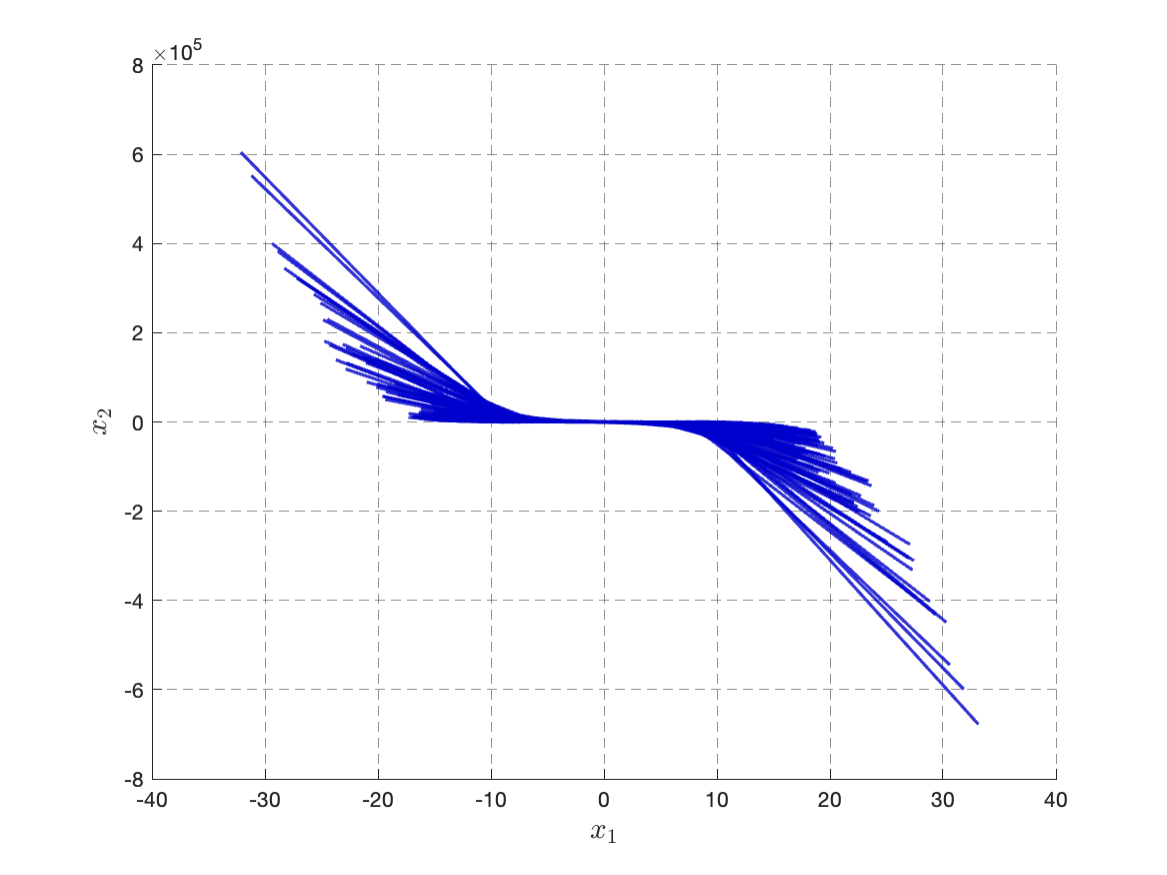}}\vspace{-0.1cm}\hspace{-0.5cm}
        \subfloat[dt-LS: DC Motor]{\includegraphics[
    width=0.34\textwidth
    ]{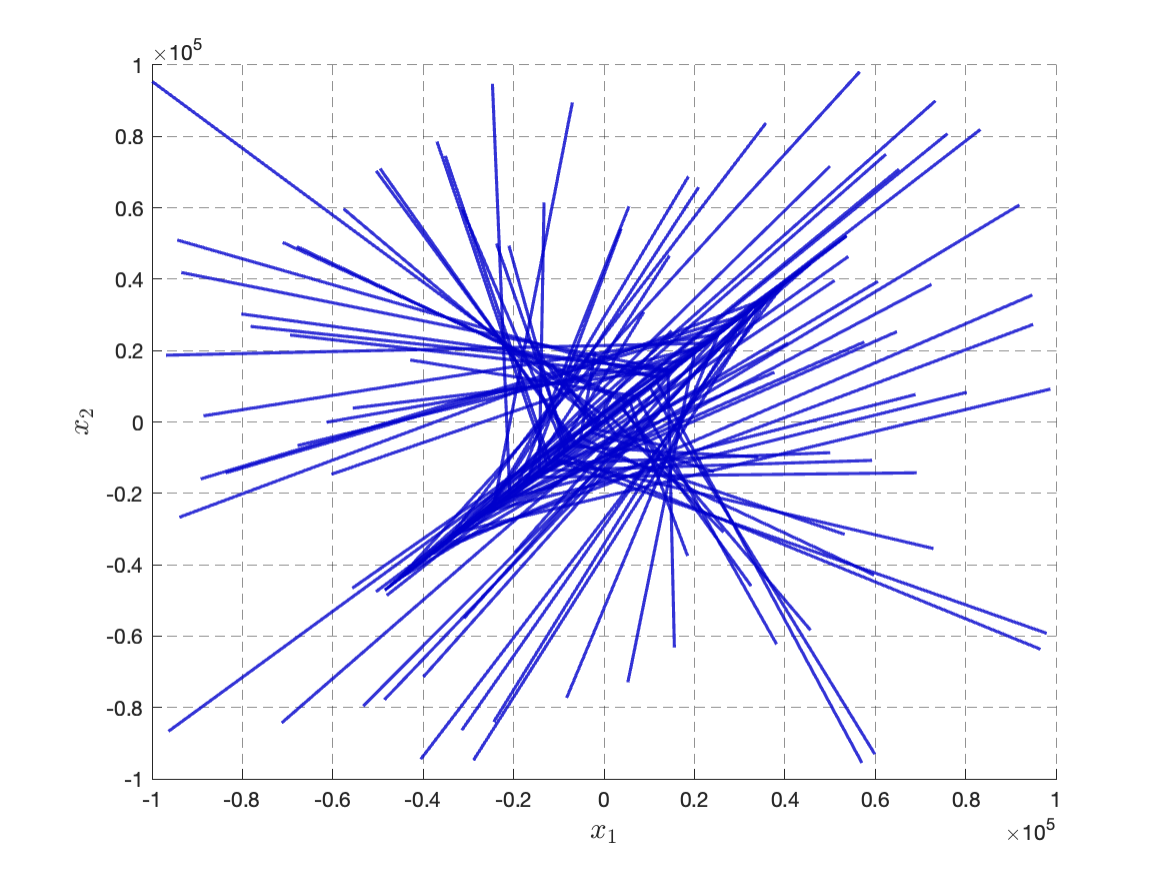}}\vspace{-0.1cm}\hspace{-0.5cm}
        \subfloat[dt-LS: Room Temperature 1]{\includegraphics[
    width=0.34\textwidth
    ]{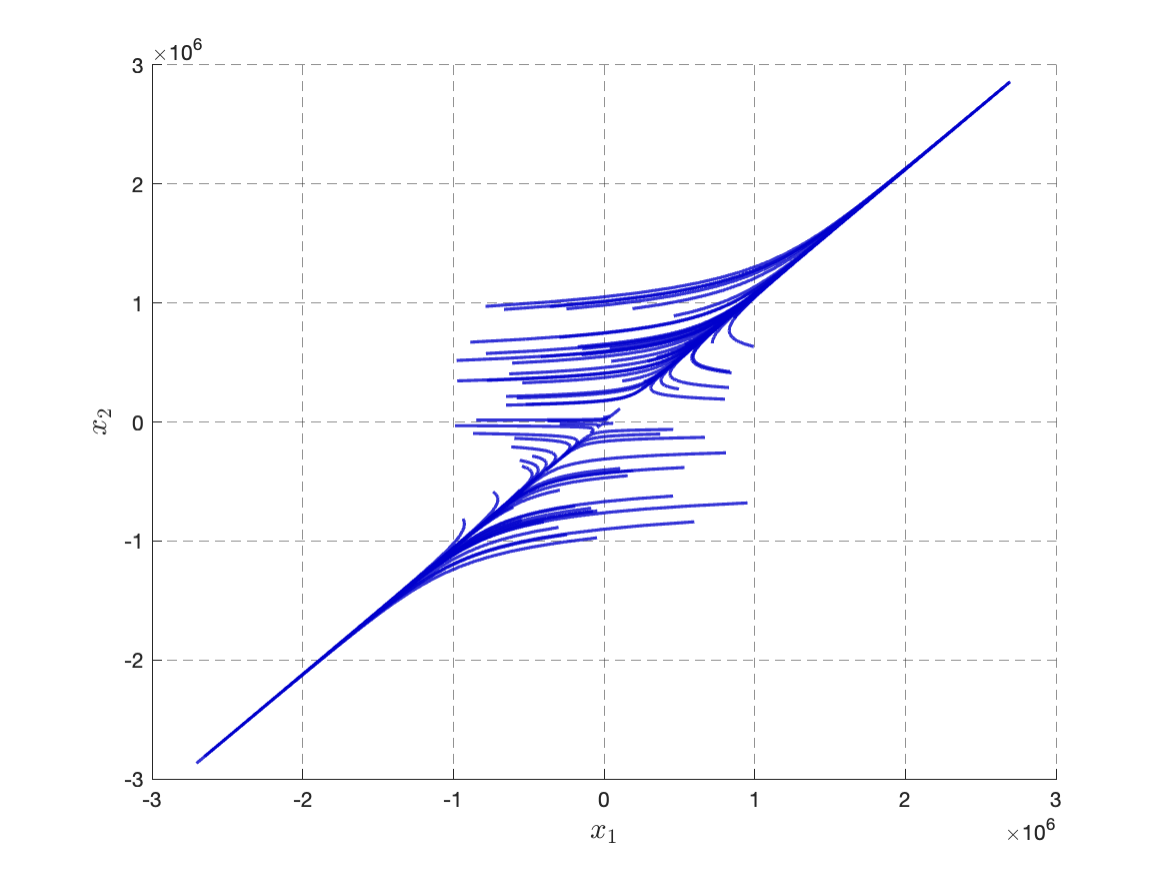}}\vspace{-0.1cm}\hspace{-0.5cm}
    \subfloat[dt-LS: Two Tank]{\includegraphics[
    width=0.34\textwidth
    ]{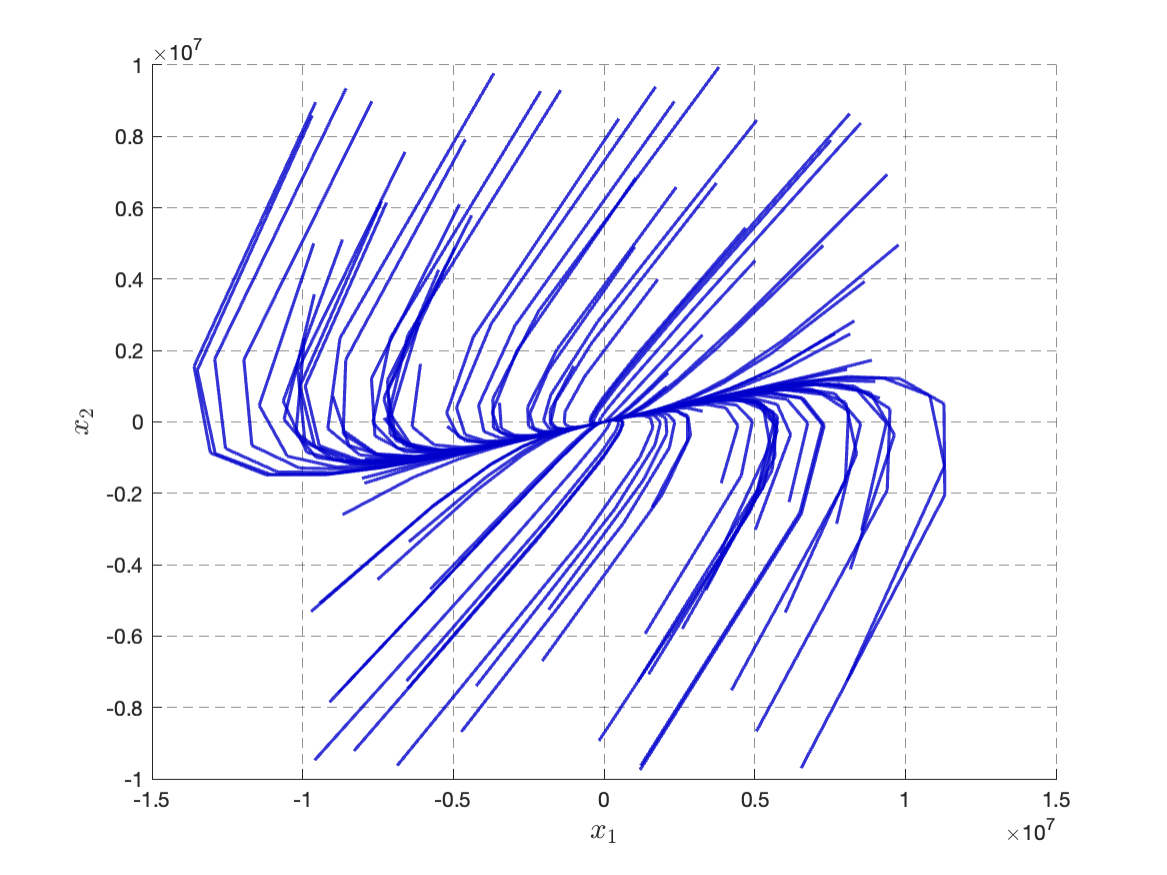}}\vspace{-0.1cm}\hspace{-0.5cm}
        \subfloat[dt-LS: Room Temperature 2]{\includegraphics[
    width=0.34\textwidth
    ]{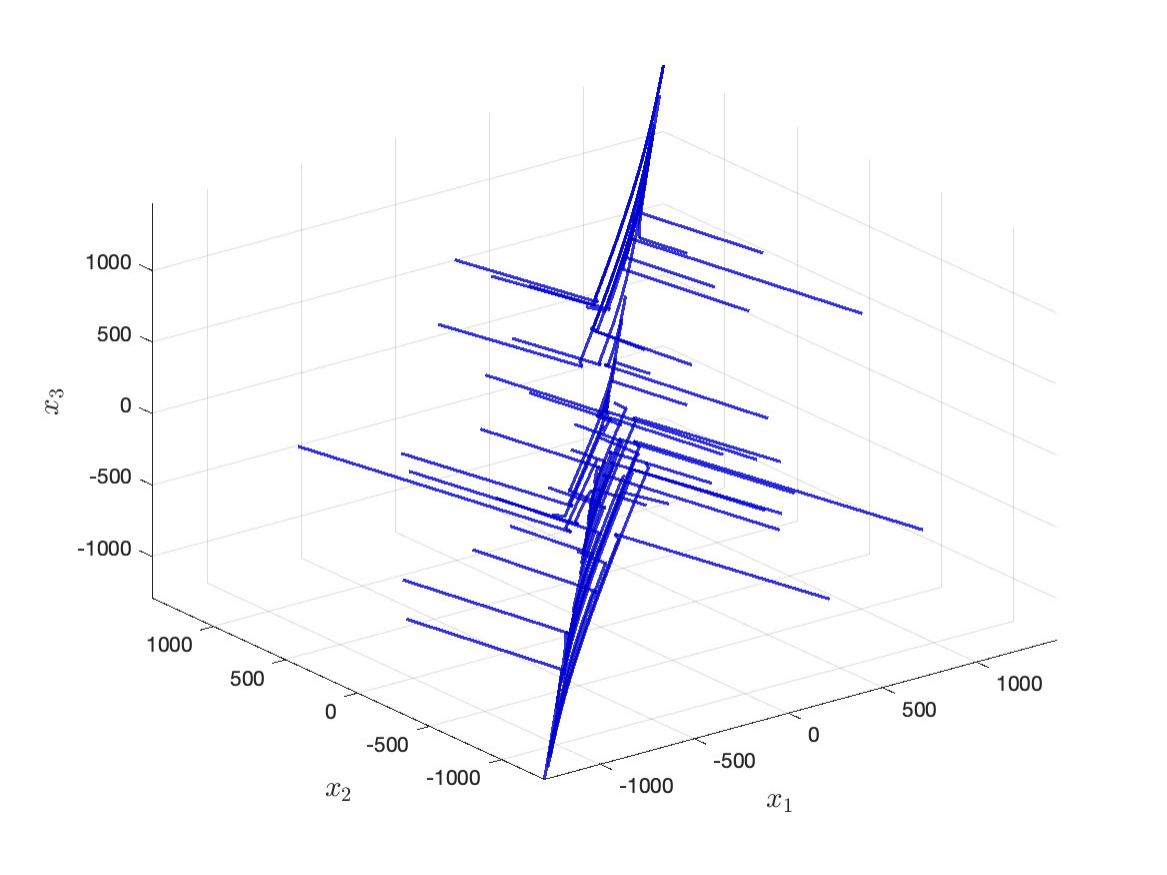}}
        \subfloat[dt-NPS: Lorenz Attractor]{\includegraphics[
    width=0.34\textwidth
    ]{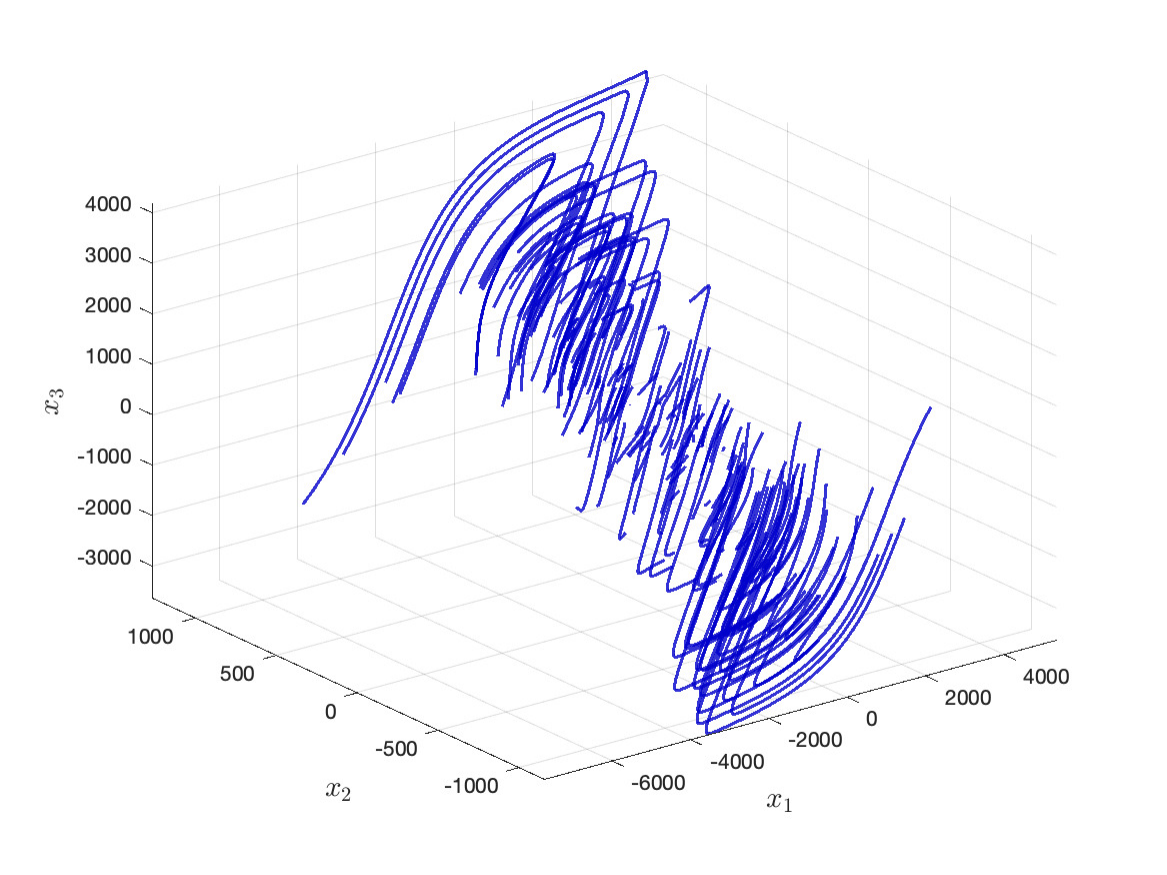}}
    \caption{\textsf{Simulation results for designing CLFs.}}
    \label{fig:CLFs}
\end{figure*}

Details of the simulations, including the number of collected samples, computation time, and memory usage, are reported in both tables. As observed, solving linear cases is very fast (under a second for 2-dimensional cases) due to the use of \emph{semidefinite programs} for satisfying the required conditions (apart from designing level sets for CBC if the property is safety-related). Nonlinear cases, however, require more computation time, as expected, due to solving \emph{SOS programs} and designing Lagrange multipliers in addition to CBC and CLF. For high-dimensional linear cases (4, 6, and 8 dimensions), solving the stability problem is pretty fast using semidefinite programs. However, the safety problem takes longer due to the need to design level sets for the CBC using an SOS optimization program. Simulation results for different case studies, illustrating the design of CBCs and CLFs, are presented in Figures~\ref{fig:CBCs} and \ref{fig:CLFs}, respectively. 
Another observation is that, although both ct-NPS models—the Lotka-Volterra Predator-Prey Model and the Van der Pol Oscillator—have a dimension of two, the computational time and memory usage for the latter are greater due to having higher-degree monomial terms. A future update to \textsf{TRUST} could include optimizing certain SOS algorithms to speed up the tool for solving nonlinear cases with potentially higher-order monomials. Additionally, whilst the tool is currently designed for scenarios without noise, future iterations of the tool will increase the scope to handle noisy data.

\begin{table*}[t!]
	\centering
	\caption{
		Data-driven design of CBCs and safety controllers.
		The symbol $n$ denotes the dimension, while $T$ represents the number of collected samples. All cases were run on a MacBook Pro (Apple M3 Max with 36 GB RAM).\vspace{-0.2cm}
	}
	{\small 	\begin{tabular}{l|c|c|c|c|c|c|c}
			\textbf{Experiment Name} & \textbf{System} & \textbf{$n$} & \textbf{$T$} & \textbf{$\gamma$} & \textbf{$\lambda$} & \textbf{Time (s)} & \textbf{Memory (MB)} \\
			\hline
			Lotka-Volterra Predator-Prey Model  & ct-NPS & 2 & 12     & 0.11 & 0.14 & 50.92 & 45.0 \\
			Van der Pol Oscillator              & ct-NPS & 2 & 15     & 5.73 & 14.26 & 345.79 & 105.5 \\
			DC Motor                            & ct-LS  & 2 & 15    & 3.11 & 3.43 & 0.53 & 1.2 \\
			Room Temperature System  1           & ct-LS  & 2 & 15    & 631.02 & 1838.62 & 0.34 & 0.8 \\
			Two Tank System                     & ct-LS  & 2 & 12    & 5.48 & 6.17 & 0.35 & 10.3\\
			High Order 4                        & ct-LS  & 4 & 16 & 13,910.89 & 14,374.75 & 5.98 & 6.2 \\
			High Order 6                        & ct-LS  & 6 & 16 & 20,072.83 & 20,600.83 & 91.06 & 17.0 \\
			High Order 8                        & ct-LS  & 8 & 20 & 47,437.80 & 66,261.32 & 1008.80 & 44.0 \\
			Lotka-Volterra Predator Prey        & dt-NPS & 2 & 12    & 0.46 & 0.57 & 59.86 & 58.3 \\
			Lorenz Attractor                    & dt-NPS & 3 & 12    & 1,052.68 & 3,931.40 & 948.52 & 323.7 \\
			DC Motor                            & dt-LS  & 2 & 15    & 0.64 & 0.70 & 0.61 & 1.4 \\
			Room Temperature System 1            & dt-LS  & 2 & 15    & 8.87 & 9.07 & 0.77 & 1.8\\
			Room Temperature System  2        & dt-LS  & 3 & 15    & 8.02 & 12.00 & 2.58 & 9.5 \\
			Two Tank System                     & dt-LS  & 2 & 8     & 1.80 & 3.17 & 1.05 & 2.3 \\
			High Order 4                        & dt-LS  & 4 & 16 & 261.63 & 264.94 & 5.59 & 5.6 \\
			High Order 6                        & dt-LS  & 6 & 16 & 22.50 & 23.14 & 85.01 & 16.0 \\
			High Order 8                        & dt-LS  & 8 & 20 & 21.93 & 30.46 & 1003.39& 48.5 \\
	\end{tabular}}
	\label{tab:safety_stable}
\end{table*}

\begin{table*}[t!]
	\centering
	\caption{
		Data-driven design of CLFs and stability controllers. The symbol $n$ denotes the dimension, while $T$ represents the number of collected samples. All cases were run on a MacBook Pro (Apple M3 Max with 36 GB RAM).\vspace{-0.2cm}
	}
	{\small 	\begin{tabular}{l|c|c|c|c|c}
			\textbf{Experiment Name} & \textbf{System} & \textbf{$n$} & \textbf{$T$} & \textbf{Time (s)} & \textbf{Memory (MB)} \\
			\hline
			Lotka-Volterra Predator-Prey Model  & ct-NPS & 2 & 12   & 49.76 & 33.8 \\
			Van der Pol Oscillator              & ct-NPS & 2 & 15   & 347.67 & 139.7 \\
			DC Motor                            & ct-LS  & 2 & 15   & 0.02 & 0.1 \\
			Room Temperature System 1            & ct-LS  & 2 & 15   & 0.05 & 0.1 \\
			Two Tank System                     & ct-LS  & 2 & 12   & 0.06 & 0.1 \\
			High Order 4                        & ct-LS  & 4 & 16    & 0.10 & 0.2 \\
			High Order 6                        & ct-LS  & 6 & 16    & 0.17 & 0.4 \\
			High Order 8                        & ct-LS  & 8 & 16    & 0.35 & 0.8 \\
			Academic System                     & dt-NPS & 2 & 12   & 44.72 & 43.9 \\
			Lorenz Attractor                    & dt-NPS & 3 & 12   & 720.81 & 251 \\
			DC Motor                            & dt-LS  & 2 & 15   & 0.06 & 0.2\\
			Room Temperature System 1          & dt-LS  & 2 & 15   & 0.05 & 0.1\\
			Room Temperature System 2 & dt-LS  & 3 & 15   & 0.08 & 0.2 \\
			Two Tank System                     & dt-LS  & 2 & 8    & 0.06 & 0.1 \\
			High Order 4                        & dt-LS  & 4 & 16   & 0.15 & 0.3\\
			High Order 6                        & dt-LS  & 6 & 16   & 0.29 & 0.6 \\
			High Order 8                        & dt-LS  & 8 & 16   & 0.49 & 1.2 \\
	\end{tabular}}
	\label{tab:stability_stable}
\end{table*}

\section{Conclusion}
We developed an open-source software tool, \textsf{TRUST}, for \emph{data-driven controller synthesis} of dynamical systems with \emph{unknown} mathematical models, to guarantee stability or safety properties. Using only a \emph{single input-state trajectory} from the unknown system and by meeting a rank condition for \textsf{persistent excitation}, \textsf{TRUST} is designed to construct either \emph{control Lyapunov functions} or \emph{control barrier certificates}, along with corresponding stability or safety controllers. The tool employs SOS optimization programs based solely on data to enforce these properties across four system classes: \emph{(i) continuous-time nonlinear polynomial systems, (ii) continuous-time linear systems, (iii) discrete-time nonlinear polynomial systems, and (iv) discrete-time linear systems}. We applied \textsf{TRUST} to a set of physical benchmarks with unknown dynamics, ensuring their stability or safety properties within the supported model classes. Future directions involve extending  \textsf{TRUST} to support a \emph{wider range of nonlinear systems} and to accommodate unknown systems with \emph{stochastic dynamics}.

\begin{algorithm}[t!]
	\caption{Data-driven design of \emph{CLF and stability} controller for \emph{dt-LS}}\label{alg:dtls-stab}
	\begin{algorithmic}[1]
		\Require Collected trajectories $\mathcal{U}_0^d,\mathcal{X}_0^d,\mathcal{X}_1^d$
		\State Check that the full row-rank condition for $\mathcal{X}_0^d$ is satisfied
		\State Solve~\eqref{eq:schur-stability-dtLS0} and~\eqref{eq:schur-stability-dtLS1} for $P$ and $H$, simultaneously
		\Ensure CLF $\mathcal{V}(x) = x^\top [\mathcal{X}_{0}^dH]^{-1}x$ and its corresponding stability controller $u = \mathcal{U}_{0}^dH[\mathcal{X}_{0}^dH]^{-1}x$
	\end{algorithmic}
\end{algorithm}

\section{Acknowledgment}
The authors would like to thank the MOSEK Team for their support, which allowed the use of the academic MOSEK license as part of the web application for our tool. Additionally, the authors are grateful to Chenyang Yuan for his assistance with the SOS toolbox~\cite{Yuan_SumOfSquares_py}. Ben Wooding is supported by an EPSRC Doctoral Prize Research Fellowship.

\bibliographystyle{ieeetran}
\bibliography{biblio}

\section{Appendix}

The mathematical models for each case study are provided below. Note that these models are assumed to be unknown, and our analysis relies solely on the trajectories collected from them.

\noindent\textbf{ct-NPS: Lotka-Volterra Predator-Prey Model.}
The Lotka-Volterra equations are to describe the dynamics of biological systems in which two species interact \cite{wangersky1978lotka}. The state variable $x_1$ represents the population density of the prey and $x_2$ the population density of the predator:
\begin{equation*}
	\Sigma^c: \begin{cases}
		\dot{x}_1(t) = \alpha x_1(t) - \beta x_1(t)x_2(t)-u_1(t),\\
		\dot{x}_2(t) = -\eta x_2(t) + \delta x_1(t)x_2(t) + u_2(t),
	\end{cases}
\end{equation*}
which is of the form~\eqref{eq:ct-NPS} with
\begin{equation*}
	A = \begin{bmatrix}
		\alpha & 0 & -\beta \\
		0 & -\eta & \delta
	\end{bmatrix}\!,\quad B = \begin{bmatrix}
		-1 & 0 \\
		0 & 1
	\end{bmatrix}\!,\quad \mathcal{M}(x) = \begin{bmatrix}
		x_1(t) \\
		x_2(t) \\
		x_1(t)x_2(t)
	\end{bmatrix}\!,
\end{equation*}
where $\alpha=\eta=0.6$  are the prey growth and predator death rates, while $\beta=\delta=1$ are the effect of the presence of the predator and prey on each other. The inputs above represent a pest control scenario where either pesticides or the introduction of predators is used to manage the prey population. For safety problems, the regions of interest are given as follows: 
state space $X=[-2, 2]\times[-1, 1]$, initial set $X_I=[0.5, 1]\times[0.2, 0.4]$ and unsafe sets $X_O=[-2, -1.5]\times[0.8, 1] \cup [-2, -1.5]\times[-1, -0.8] \cup [1.6, 2]\times[0.85, 1] \cup [1.6, 2]\times[-1, -0.8]$.

\noindent\textbf{ct-NPS: Van der Pol Oscillator.}
We consider the Van der Pol oscillator benchmark from the ARCH competition~\cite{abate2020arch}, with the following dynamics:
\begin{equation*}
    \Sigma^c\!:\begin{cases}
        \dot{x}_1(t) = x_2(t),\\
        \dot{x}_2(t) = -x_1(t) + (1-x_1(t)^2)x_2(t) + u(t),
    \end{cases}
\end{equation*}
which is of the form~\eqref{eq:ct-NPS} with
\begin{equation*}
	A = \begin{bmatrix}
		0 & 1 & 0 \\
		-1 & 1 & -1
	\end{bmatrix}\!,\quad B = \begin{bmatrix}
		0 & 1
	\end{bmatrix}^\top\!,\quad \mathcal{M}(x) = \begin{bmatrix}
		x_1(t) \\
		x_2(t) \\
		x_1^2(t)x_2(t)
	\end{bmatrix}\!.
\end{equation*}
For safety problems, we consider $X = [-2,2]^2$, $X_I=[-0.2,0.2]^2$ and
$X_O=[-2,-1.5]^2\cup[1.5,2]^2$.

\noindent\textbf{ct-LS: DC Motor.} We consider a DC Motor, based on~\cite{adewuyi2013dc}, with the dynamics
\begin{align}\label{new18}
	\Sigma^c : \begin{cases}
		\dot x_1 =- (\frac{R}{L}x_1 + \frac{k_{dc}}{L}x_2 + u_1), \\
		\dot x_2 = \frac{k_{dc}}{J}x_1 - \frac{b}{J}x_2 + u_2,
	\end{cases}
\end{align}
which is of the form~\eqref{eq:ct-LS} with
\begin{align*}
	A = \begin{bmatrix}
		-\frac{ R}{L} & -\frac{ k_{dc}}{L} \\
		\frac{ k_{dc}}{J} & - \frac{ b}{J}  \\
	\end{bmatrix}\!,\quad B = \begin{bmatrix}
		1 & 0  \\
		0 & 1 \\
	\end{bmatrix}\!,
\end{align*}
where $x_1,x_2,R=1,L=0.01,J=0.01$ are the armature current, the rotational speed of the shaft, the electrical resistance, the electrical inductance, and the moment of inertia of the rotor, respectively. In addition, $b=1$, and $k_{dc}=0.01$, represent, respectively, the motor torque, and the back electromotive force. For safety problems, the state space is given as $X=[-1, 1]^2$, with initial set $X_I=[0.1, 0.4]\times[0.1, 0.55]$ and
four unsafe sets $X_O=[0.5, 1]\times[0.6, 1] \cup [-1, -0.6]\times[0.6, 1] \cup [-1, -0.7]\times[-1, -0.6] \cup [0.6, 1]\times[-1, -0.6]$.

\noindent\textbf{ct-LS: Room Temperature System 1.} We consider the two-room system~\cite{abate2021fossil}, with dynamics
\begin{equation}\label{new19}
	\Sigma^c:\begin{cases}
		\dot x_1  = - (\alpha + \alpha_{e1})x_1 + \alpha x_2 + \alpha_{e1} u,\\
		\dot x_2 =  -(\alpha + \alpha_{e2})x_2 + \alpha x_1 + \alpha_{e2}u,
	\end{cases}
\end{equation}
which is of the form~\eqref{eq:ct-LS} with
\begin{align*}
	A = \begin{bmatrix}
		- (\alpha + \alpha_{e1}) & \alpha \\
		\alpha  &	- (\alpha + \alpha_{e2})\\
	\end{bmatrix}\!,\quad B = \begin{bmatrix}
		\alpha_{e1}  \\
		\alpha_{e2} \\
	\end{bmatrix}\!,
\end{align*}
where heat exchange constants are $\alpha = 5\times 10^{-2}, \alpha_{e1} = 5\times 10^{-3}, \alpha_{e2} = 8\times10^{-3}$. For safety problems, the state space is provided as $X=[-2, 3]^2$, with the initial set $X_I=[-0.5, 0.5]^2$ and two unsafe sets $X_O=[-2,-1]\times[2, 3] \cup [2, 3]\times[-2, -1]$.

\noindent\textbf{ct-LS: Two Tank System.} Consider a two-tank system~\cite{ramos2007mathematical}, characterized by differential equations
\begin{equation}\label{new20}
	\Sigma^c:\begin{cases}
		\dot x_1 = -\frac{\alpha_1}{a_1}x_1 + \frac{u_1}{a_1},\\
		\dot x_2 = \frac{\alpha_1}{a_2}x_1 -\frac{\alpha_2}{a_2}x_2 + \frac{u_2}{a_2},
	\end{cases}
\end{equation}
which is of the form~\eqref{eq:ct-LS} with
\begin{align*}
	A = \begin{bmatrix}
		-\frac{\alpha_1}{a_1} & 0 \\
		\frac{\alpha_1}{a_2} & -\frac{\alpha_2}{a_2} \\
	\end{bmatrix}\!,\quad B = \begin{bmatrix}
		\frac{1}{a_1} & 0 \\
		0 & \frac{1}{a_2} \\
	\end{bmatrix}\!,
\end{align*}
where $x_1$, $x_2$ are heights of the fluid in two tanks. Additionally, $\alpha_i$ and $a_i$ are the valve coefficient and area of the tank $i$, and $u_1$ and $u_2$ are the inflow and outflow rate of tank $1$ and $2$, respectively. Furthermore,  $\alpha_1 =\alpha_2=a_1=a_2=2$. For safety problems, we consider the state space $X=[-3, 3]^2$, initial set $X_I=[-1, 1]^2$ and two unsafe sets 
$X_O=[1.5, 3]^2 \cup [-3, -1.5]^2$.

\noindent\textbf{ct-LS: 4D Academic Example.}
We consider the following $4$-dimensional benchmark adapted from~\cite{abate2021fossil}
\begin{equation*}
	\Sigma^c:\begin{cases}
		\dot{x}_1(t) = x_2(t),\\
		\dot{x}_2(t) = x_3(t),\\
		\dot{x}_3(t) = x_4(t),\\
		\dot{x}_4(t) = -3980x_4(t) - 4180x_3(t) - 2400x_2(t) - 576x_1(t) + u(t),
	\end{cases}
\end{equation*}
which is of the form~\eqref{eq:ct-LS} with
\begin{align*}
	A &= \begin{bmatrix}
		0 & 1 & 0 & 0\\
		0 & 0 & 1 & 0 \\
		0 & 0 & 0 & 1 \\
		-576 & -2400 & -4180 & -3980
	\end{bmatrix}\!,\\\quad B &= \begin{bmatrix}
		0 & 0&0&0&0&1
	\end{bmatrix}^\top\!\!,
\end{align*}
with the state space $X=[-2,2]^4$, initial region $X_I=[0.5,1.5]^4$, and unsafe region $X_O=[-2.4,-1.6]^4$ for safety problems.

\noindent\textbf{ct-LS: 6D Academic Example.}
We consider the following $6$-dimensional benchmark adapted from~\cite{abate2021fossil}
\begin{equation*}
	\Sigma^c:\begin{cases}
		\dot{x}_1(t) = &x_2(t),\\
		\dot{x}_2(t) = &x_3(t),\\
		\dot{x}_3(t) = &x_4(t),\\
		\dot{x}_4(t) = &x_5(t),\\
		\dot{x}_5(t) = &x_6(t),\\
		\dot{x}_6(t) = &-800x_6(t) - 2273x_5(t) -3980x_4(t) - 4180x_3(t) \\&- 2400x_2(t) - 576x_1(t) + u(t),
	\end{cases}
\end{equation*}
which is of the form~\eqref{eq:ct-LS} with
\begin{align*}
	A &= \begin{bmatrix}
		0 & 1 & 0 & 0 & 0 & 0\\
		0 & 0 & 1 & 0 & 0 & 0 \\
		0 & 0 & 0 & 1 & 0 & 0 \\
		0 & 0 & 0 & 0 & 1 & 0 \\
		0 & 0 & 0 & 0 & 0 & 1 \\
		-576 & -2400 & -4180 & -3980 & -2273 & -800
	\end{bmatrix}\!,\\\quad B &= \begin{bmatrix}
		0 & 0&0&0&0&1
	\end{bmatrix}^\top\!,
\end{align*}
with the state space $X=[-2,2]^6$, initial region $X_I=[0.5,1.5]^6$, and unsafe region $X_O=[-2,-1.6]^6$ for safety problems.

\noindent\textbf{ct-LS: 8D Academic Example.}
We consider the following $8$-dimensional benchmark adapted from~\cite{abate2021fossil}
\begin{equation*}
    \Sigma^c:\begin{cases}
        \dot{x}_1(t) = &x_2(t),\\
        \dot{x}_2(t) = &x_3(t),\\
        \dot{x}_3(t) = &x_4(t),\\
        \dot{x}_4(t) = &x_5(t),\\
        \dot{x}_5(t) = &x_6(t),\\
        \dot{x}_6(t) = &x_7(t),\\
        \dot{x}_7(t) = &x_8(t),\\
        \dot{x}_8(t) = &-20x_8(t)-170x_7(t)-800x_6(t) - 2273x_5(t)\\ &-3980x_4(t) - 4180x_3(t) - 2400x_2(t) - 576x_1(t) + u(t),
    \end{cases}
\end{equation*}
which is of the form~\eqref{eq:ct-LS} with
\begin{align*}
	A &\!=\! \begin{bmatrix}
		 0 & 1 & 0 & 0 & 0 & 0 & 0 & 0\\
   0 & 0 & 1 & 0 & 0 & 0 & 0 & 0\\
   0 & 0 & 0 & 1 & 0 & 0 & 0 & 0\\
   0 & 0 & 0 & 0 & 1 & 0 & 0 & 0\\
   0 & 0 & 0 & 0 & 0 & 1 & 0 & 0\\
   0 & 0 & 0 & 0 & 0 & 0 & 1 & 0\\
   0 & 0 & 0 & 0 & 0 & 0 & 0 & 1\\
   -576 & -2400 & -4180 & -3980 & -2273 & -800 & -170 & -20
	\end{bmatrix}\!,\\\quad B &= \begin{bmatrix}
	   0 & 0&0&0&0&0&0&1
	\end{bmatrix}^\top\!,
\end{align*}
with the state space $X=[-2.2,2.2]^8$, initial region $X_I=[0.9,1.1]^8$, and unsafe region $X_O=[-2.2,-1.8]^8$ for safety problems.

\noindent\textbf{dt-NPS: Lotka-Volterra Predator-Prey Model.} We consider the Lotka-Volterra Predator-Prey Model \cite{wangersky1978lotka}, with dynamics

\begin{equation*}
	\Sigma^d: \begin{cases}
		x_1^+ = x_1 + \tau (\alpha x_1 - \beta x_1x_2 - u_1),\\
		x_2^+  = x_2 + \tau (-\eta x_2 + \delta x_1x_2+ u_2),
	\end{cases}
\end{equation*}
which is of the form~\eqref{eq:dt-NPS} with
\begin{align*}
	A = \begin{bmatrix}
		1+ \tau\alpha & 0 & -\tau\beta\\
		0 & 1- \tau\eta & \tau\delta \\
	\end{bmatrix}\!,\quad B = \begin{bmatrix}
		-\tau  \\
		\tau
	\end{bmatrix}\!, \quad\mathcal{M}(x) = \begin{bmatrix}
		x_1 \\
		x_2 \\
		x_1x_2
	\end{bmatrix}\!,
\end{align*}
where $\tau=0.01,\alpha=0.6,\beta=1.0,\eta=0.6,\delta=1.0$. 
For safety problems, we consider $X = [-2,2]\times[-1,1]$, $\initial=[0.5,1]\times[0.2,0.4]$, and $\obstruction=[-2,-1.5]\times[0.8,1]\cup[-2,-1.5]\times[-1,-0.8]\cup[1.6,2]\times[0.85,1]\cup[1.6,2]\times[-1,-0.8]$.

\noindent\textbf{dt-NPS: Academic System.} We focus on the following nonlinear polynomial system borrowed from~\citep{guo2020learning}:
\begin{align}\notag
	\Sigma^d: \begin{cases}
		x_1^+  = x_1 + \tau x_2,\\\label{case_study}
	x_2^+  = x_2 + \tau (x_1^2+u),
		\end{cases}
\end{align}
which is of the form of \eqref{eq:dt-NPS}, with
\begin{align*}
	A = \begin{bmatrix}1&0\\0&1\end{bmatrix}\!, \quad B = \begin{bmatrix}0\\1\end{bmatrix}\!, \quad \mathcal M(x) = \begin{bmatrix}x_2\\x_1^2\end{bmatrix}\!.
\end{align*}
with $\tau = 0.1$. Note that this case study serves solely for stability; therefore, no regions of interest for safety are provided.

\noindent\textbf{dt-NPS: Lorenz Attractor.}
We consider the Lorenz attractor, a well-studied dynamical system with chaotic behavior, with dynamics
\begin{equation*}
	\Sigma^d : \begin{cases}
		x_1^+  = x_1+ \tau (\sigma x_1 + \sigma x_2 + u_1), \\
		x_2^+  = x_2 \!+\!\tau (\rho x_1 \!-\! x_2 \!-\! x_1x_3  \!+\! u_2), \\
		x_3^+  = x_3 + \tau (x_1x_2-\beta x_3+u_3),
	\end{cases}
\end{equation*}
which is of the form~\eqref{eq:dt-NPS} with
\begin{align*}
	A = \begin{bmatrix}
		1+ \tau\sigma & \tau\sigma & 0 & 0 & 0 & 0\\
		\tau\rho & 1-\tau & 0 & -\tau & 0 & 0 \\
		0 & 0 & 1-\tau\beta & \tau & 0 & 0 \\
	\end{bmatrix}\!,\\\quad B = \begin{bmatrix}
		\tau  \\
		\tau \\
		\tau
	\end{bmatrix}\!, \quad\mathcal{M}(x) = \begin{bmatrix}
		x_1\\
		x_2\\x_3\\x_1x_2\\x_2x_3\\x_1x_3
	\end{bmatrix}\!,
\end{align*}
where $\rho=28, \sigma=10,\beta=\frac{8}{3},\tau=10^{-3}$. For safety problems, the regions of interest are given as: state space $X=[-5,5]^3$, initial set $X_I=[-1, 1]^3$ and two unsafe sets
$X_O=[-5,-3.5]^3\cup[3.5,5]^3$.

\noindent\textbf{dt-LS: DC Motor.} We consider a DC Motor, based on~\cite{adewuyi2013dc}, with the dynamics
\begin{align*}
	\Sigma^d : \begin{cases}
		x_1^+  = x_1 - \tau(\frac{R}{L}x_1+ \frac{k_{dc}}{L}x_2+ u_1), \\
		x_2^+  = x_2 + \tau(\frac{k_{dc}}{J}x_1 - \frac{b}{J}x_2 + u_2),
	\end{cases}
\end{align*}
which is of the form~\eqref{eq:dt-LS} with
\begin{align*}
	A = \begin{bmatrix}
		1 -\frac{\tau R}{L} & -\frac{\tau k_{dc}}{L} \\
		\frac{\tau k_{dc}}{J} & 1 - \frac{\tau b}{J}  \\
	\end{bmatrix}\!,\quad B = \begin{bmatrix}
		1 & 0  \\
		0 & 1 \\
	\end{bmatrix}\!,
\end{align*}
where $\tau=0.01$ is the sampling time, and other parameters are as in~\eqref{new18}. For safety problems, the state space is given as $X=[-1, 1]^2$, with initial set $X_I=[0.1, 0.4]\times[0.1, 0.55]$ and two unsafe sets $X_O=[0.45, 1]\times[0.6, 1]\cup[-1, -0.6]\times[0.6, 1]$.

\noindent\textbf{dt-LS: Room Temperature System 1.} We consider the two-room system~\cite{abate2021fossil}, with dynamics
\begin{equation*}
	\Sigma^d:\begin{cases}
		x_1^+  = (1 - \tau(\alpha + \alpha_{e1}))x_1 + \tau\alpha x_2 + \tau\alpha_{e1} u,\\
		x_2^+  = (1 - \tau(\alpha + \alpha_{e2}))x_2 + \tau\alpha x_1 + \tau\alpha_{e2}u,
	\end{cases}
\end{equation*}
which is of the form~\eqref{eq:dt-LS} with
\begin{align*}
	A = \begin{bmatrix}
		1 - \tau(\alpha + \alpha_{e1}) & \tau\alpha \\
		\tau\alpha & 1 - \tau(\alpha + \alpha_{e2}) \\
	\end{bmatrix}\!,\quad B = \begin{bmatrix}
		\tau\alpha_{e1}  \\
		\tau\alpha_{e2} \\
	\end{bmatrix}\!,
\end{align*}
where $\tau = 5$, and other parameters are as in~\eqref{new19}. For safety problems,
the regions of interest as given as $X=[-2,3]^2$, $X_I=[-0.5, 0.5]^2$, and $X_O=[2, 3]^2 \cup [-2, -0.5]\times[1.5, 3] \cup [1.5, 3]\times[-2, -0.5]$.

\noindent\textbf{dt-LS: Room Temperature System 2.}  We consider the three-room system~\cite{abate2021fossil}, with dynamics

\begin{equation*}
	\Sigma^d:\begin{cases}
		x_1^+ = (1 - \tau(\alpha - \alpha_{e1}))x_1 + \tau\alpha x_3 + \tau\alpha_{e1} u,\\
		x_2^+  = (1 - \tau(\alpha - \alpha_{e2}))x_2 + \tau\alpha (x_1 + x_3)+ \tau\alpha_{e2}u,\\
            x_3^+ = (1 - \tau(\alpha - \alpha_{e3}))x_3 + \tau\alpha x_1 + \tau\alpha_{e3}u,
	\end{cases}
\end{equation*}
which is of the form~\eqref{eq:dt-LS} with
\begin{align*}
	A = \begin{bmatrix}
		1 \!-\! \tau(\alpha \!-\! \alpha_{e1}) & 0 & \tau\alpha \\
		\tau\alpha & 1 \!-\! \tau(2\alpha \!-\! \alpha_{e2}) & \tau\alpha \\
		\tau\alpha & 0 & 1 \!-\! \tau(\alpha \!-\! \alpha_{e3})
	\end{bmatrix}\!,~ B = \begin{bmatrix}
		\tau\alpha_{e1}  \\
		\tau\alpha_{e2} \\
            \tau\alpha_{e3} \\
	\end{bmatrix}\!,
\end{align*}
where
$\tau = 5$, and other parameters are as in~\eqref{new19}.
For safety problems, the regions of interest are given as: state space $X=[-2, 3]^3$, initial set $X_I=[-0.5, 0.5]^3$ and two unsafe sets $X_O=[2, 3]\times[-2, -3]\times[2, 3] \cup [-2, -3]\times[2, 3]\times[-2, -3]^3$.

\noindent\textbf{dt-LS: Two Tank System.} Consider a two-tank system~\cite{ramos2007mathematical}, characterized by difference equations
\begin{equation*}
	\Sigma^d:\begin{cases}
		x_1^+  = (1-\tau\frac{\alpha_1}{a_1})x_1 + \tau\frac{u_1}{a_1},\\
		x_2^+  = \tau\frac{\alpha_1}{a_2}x_1 + (1-\tau\frac{\alpha_2}{a_2})x_2+ \tau\frac{u_2}{a_2},
	\end{cases}
\end{equation*}
which is of the form~\eqref{eq:dt-LS} with
\begin{align*}
	A = \begin{bmatrix}
		1-\tau\frac{\alpha_1}{a_1} & 0 \\
		\tau\frac{\alpha_1}{a_2} & 1-\tau\frac{\alpha_2}{a_2} \\
	\end{bmatrix}\!,\quad B = \begin{bmatrix}
		\tau\frac{1}{a_1} & 0 \\
		0 & \tau\frac{1}{a_2} \\
	\end{bmatrix}\!,
\end{align*}
where $\tau=0.1$, and other parameters are as in~\eqref{new20}. 
For safety problems, the regions of interest are $X=[-2, 2]^2$, $\initial=[-0.5, 0.5]^2$, $\obstruction=[1.5, 2]^2 \cup [-2, -1.5] \times [1, 2] \cup [-1.5, -1] \times [1.5, 2] \cup [1.5, 2] \times [-2, -1]$.

\noindent\textbf{dt-LS: 4D Academic Example.}
We consider the following $4$-dimensional benchmark adapted from~\cite{abate2021fossil}
\begin{equation*}
	\Sigma^d:\begin{cases}
		x_1^+  = x_1 + \tau(x_2+u_1),\\
		x_2^+  = x_2 + \tau(x_3+u_2),\\
		x_3^+  = x_3+ \tau(x_4+u_3),\\
		x_4^+  = x_4 + \tau(-3980x_4 - 4180x_3 - 2400x_2 - 576x_1 + u_4),
	\end{cases}
\end{equation*}
which is of the form~\eqref{eq:dt-LS} with
\begin{align*}
	A &= \begin{bmatrix}
		1 & \tau & 0 & 0\\
		0 & 1 & \tau & 0 \\
		0 & 0 & 1 & \tau \\
		-576\tau & -2400\tau & -4180\tau & 1-3980\tau
	\end{bmatrix}\!,\\\quad B &= \tau~\!\identity_4,
\end{align*}
where $\tau=0.001$, with the state space $X=[-2,2]^4$, initial region $X_I=[0.5,1.5]^4$, and unsafe region $X_O=[-2.4,-1.6]^4$ for safety problems.

\noindent\textbf{dt-LS: 6D Academic Example.}
We consider the following $6$-dimensional benchmark adapted from~\cite{abate2021fossil}
\begin{equation*}
	\Sigma^d:\begin{cases}
		x_1^+  = x_1 + \tau(x_2+u_1),\\
		x_2^+  = x_2 + \tau(x_3+u_2),\\
		x_3^+  = x_3+ \tau(x_4+u_3),\\
		x_4^+  = x_4 + \tau(x_5+u_4),\\
		x_5^+  = x_5 + \tau(x_6+u_5),\\
		x_6^+  = x_6 + \tau(-800x_6 - 2273x_5 -3980x_4 - 4180x_3 - 2400x_2 \\
		\quad\quad- 576x_1 + u_6),
	\end{cases}
\end{equation*}
which is of the form~\eqref{eq:dt-LS} with
\begin{align*}
	A &= \begin{bmatrix}
		1 & \tau & 0 & 0 & 0 & 0\\
		0 & 1 & \tau & 0 & 0 & 0 \\
		0 & 0 & 1 & \tau & 0 & 0 \\
		0 & 0 & 0 & 1 & \tau & 0 \\
		0 & 0 & 0 & 0 & 1 & \tau \\
		-576\tau & -2400\tau & -4180\tau & -3980\tau & -2273\tau & 1-800\tau
	\end{bmatrix}\!,\\\quad B &= \tau~\!\identity_6,
\end{align*}
where $\tau=0.1$, with the state space $X=[-2,2]^6$, initial region $X_I=[0.5,1.5]^6$, and unsafe region
$X_O=[-2, -1.6]^6$ for safety problems.

\noindent\textbf{dt-LS: 8D Academic Example.}
We consider the following $8$-dimensional benchmark adapted from~\cite{abate2021fossil}
\begin{equation*}
    \Sigma^d:\begin{cases}
		x_1^+  = x_1+ \tau(x_2+u_1),\\
		x_2^+  = x_2 + \tau(x_3+u_2),\\
		x_3^+  = x_3 + \tau(x_4+u_3),\\
		x_4^+  = x_4 + \tau(x_5+u_4),\\
		x_5^+  = x_5 + \tau(x_6+u_5),\\
        x_6^+  = x_6 + \tau(x_7+u_6),\\
        x_7^+  = x_7 + \tau(x_8+u_7),\\
        x_8^+  = x_8 + \tau(-20x_8-170x_7-800x_6 -  2273x_5-3980x_4- 4180x_3  \\  \quad\quad- 2400x_2- 576x_1+ u_8),
    \end{cases}
\end{equation*}
which is of the form~\eqref{eq:dt-LS} with
\begin{align*}
	A &\!=\! \begin{bmatrix}
		 1 \!\!&\!\! \tau \!\!&\!\! 0 \!\!&\!\! 0 \!\!&\!\! 0 \!\!&\!\! 0 \!\!&\!\! 0 \!\!&\!\! 0\\
   0 \!\!&\!\! 1 \!\!&\!\! \tau \!\!&\!\!  0 \!\!&\!\!  0 \!\!&\!\!  0 \!\!&\!\!  0 \!\!&\!\!  0\\
   0 \!\!&\!\!  0 \!\!&\!\! 1 \!\!&\!\!  \tau \!\!&\!\!  0 \!\!&\!\!  0 \!\!&\!\!  0 \!\!&\!\!  0\\
   0 \!\!&\!\!  0 \!\!&\!\! 0 \!\!&\!\!  1 \!\!&\!\!  \tau \!\!&\!\!  0 \!\!&\!\!  0 \!\!&\!\!  0\\
   0 \!\!&\!\!  0 \!\!&\!\!  0 \!\!&\!\!  0 \!\!&\!\!  1 \!\!&\!\!  \tau \!\!&\!\!  0 \!\!&\!\!  0\\
   0 \!\!&\!\!  0 \!\!&\!\!  0 \!\!&\!\!  0 \!\!&\!\!  0 \!\!&\!\! 1 \!\!&\!\!  \tau \!\!&\!\!  0\\
   0 \!\!&\!\!  0 \!\!&\!\!  0 \!\!&\!\!  0 \!\!&\!\!  0 \!\!&\!\!  0 \!\!&\!\!  1 \!\!&\!\!  \tau\\
   -576\tau \!\!&\!\!  -2400\tau \!\!&\!\!  -4180\tau \!\!&\!\!  -3980\tau \!\!&\!\!  -2273\tau \!\!&\!\!  -800\tau \!\!&\!\!  -170\tau \!\!&\!\!  1-20\tau
	\end{bmatrix}\!,\\\quad B &= \tau~\!\identity_8,
\end{align*}
where $\tau=0.1$, with the state space $X=[-2.2,2.2]^8$, initial region $X_I=[0.9,1.1]^8$, and unsafe region $X_O=[-2.2,-1.8]^8$ for safety problems.

\end{document}